\documentclass{aa}  
\makeatletter
\renewcommand*\aa@pageof{, page \thepage{} of \pageref*{LastPage}}
\makeatother
\usepackage{graphicx}
\usepackage{amsmath}
\usepackage{txfonts}
\usepackage{hyperref}
\newcommand{\macc}{$\dot{M}_{\text{acc}}$}
\newcommand{\medd}{$\dot{M}_{\text{Edd}}$}

\usepackage{color}

\begin{document} 

 %  \title{Infrared observations of the TDE 2019azh reveal deviations from a blackbody spectrum}
   \title{Infrared observations reveal the reprocessing envelope in the tidal disruption event AT 2019azh}
   \author{Thomas M. Reynolds
          \inst{1,2,3}
          \and
          Lars Thomsen\inst{4}
          \and
          Seppo Mattila \inst{1,5}  
        \and
        Takashi Nagao \inst{1,6,7}
        \and
        Joseph P. Anderson\inst{8}
        \and
        Franz E. Bauer\inst{9}
        \and
        Panos Charalampopoulos \inst{1}
        \and
        Lixin Dai\inst{4}
        \and
        Sara Faris\inst{10}
        \and
        Mariusz Gromadzki\inst{11}
        \and
        Claudia P. Guti\'errez\inst{12,13}
        \and
        Hanin Kuncarayakti\inst{1}
        \and
        Cosimo Inserra\inst{14}
        \and
        Erkki Kankare\inst{1}
        \and
        Timo Kravtsov\inst{1}
         \and
        Shane Moran\inst{1,15}
        \and
        Phil Wiseman\inst{16}
          %\fnmsep\thanks{Just to show the usage
          %of the elements in the author field}
          }

   \institute{Tuorla observatory, Department of Physics and Astronomy, University of Turku, FI-20014 Turku, Finland 
             % \email{treynolds1729@gmail.com}
            \and
            Niels Bohr Institute, University of Copenhagen, Jagtvej 128, 2200 København N, Denmark
            \and
            Cosmic Dawn Center (DAWN)
            \and
            Department of Physics, University of Hong Kong, Pokfulam Road, Hong Kong
            \and
            School of Sciences, European University Cyprus, Diogenes Street, Engomi, 1516 Nicosia, Cyprus
            \and
            Aalto University Mets\"ahovi Radio Observatory, Mets\"ahovintie 114, 02540 Kylm\"al\"a, Finland
            \and
            Aalto University Department of Electronics and Nanoengineering, P.O. BOX 15500, FI-00076 AALTO, Finland
            \and
            European Southern Observatory, Alonso de C\'ordova 3107, Casilla 19, Santiago, Chile
            \and
             Instituto de Alta Investigaci{\'{o}}n, Universidad de Tarapac{\'{a}}, Casilla 7D, Arica, Chile
             \and
            School of Physics and Astronomy, Tel Aviv University, Tel Aviv 69978, Israel
             \and
             Astronomical Observatory, University of Warsaw, Al. Ujazdowskie 4,00-478 Warszawa, Poland
            \and
            Institut d'Estudis Espacials de Catalunya (IEEC), Edifici RDIT,Campus UPC, 08860 Castelldefels (Barcelona), Spain
            \and
            Institute of Space Sciences (ICE, CSIC), Campus UAB, Carrer de Can Magrans, s/n, E-08193 Barcelona, Spain
            \and
            Cardiff Hub for Astrophysics Research and Technology, School of Physics \& Astronomy, Cardiff University, Queens Buildings, The Parade, Cardiff, CF24 3AA, UK
            \and  
             School of Physics and Astronomy, University of Leicester, University Road, Leicester LE1 7RH, UK
             \and 
             School of Physics and Astronomy, University of Southampton, Southampton, SO17 1BJ, UK
             %\email{c.ptolemy@hipparch.uheaven.space}
             %\thanks{The university of heaven temporarily does not
             %        accept e-mails}
             }

   \date{}

  \abstract
  % context heading (optional)
  % {} leave it empty if necessary  
   {Tidal disruption events (TDEs) are expected to release much of their energy in the far-ultraviolet (UV), which we do not observe directly. However, infrared (IR) observations can observe re-radiation of the optical/UV emission from dust, and if this dust is observed in the process of sublimation, we can infer the un-observed UV radiated energy. TDEs have also been predicted to show spectra shallower than a blackbody in the IR, but this has not yet been observed.}
  % aims heading (mandatory)
   {We present near/mid-IR observations of the TDE AT~2019azh spanning from -3~d before peak until >1750~d after. We evaluate these observations for consistency with dust emission or direct emission from the TDE.}
  % methods heading (mandatory)
   {We fit the IR data with a modified blackbody associated with dust emission. The UV+optical+IR data are compared with simulated spectra produced from general relativistic radiation magnetohydrodynamics simulations of super-Eddington accretion. We model the data at later times (>200~d) as an IR echo.}
  % results heading (mandatory)
   {The IR data at the maximum light can not be self-consistently fit with dust emission. Instead, the data can be better fit with a reprocessing model, with the IR excess arising due to the absorption opacity being dominated by free-free processes in the dense reprocessing envelope. We infer a large viewing angle of $\sim$60\textdegree, consistent with previously reported X-ray observations, and a tidally disrupted star with mass > 2 M$_{\odot}$. The IR emission at later times is consistent with cool dust emission. We model these data as an IR echo and find that the dust is distant (0.65~pc), and clumpy, with a low covering factor. We show that TDEs can have an IR excess not arising from dust and that IR observations at early times can constrain the viewing angle for the TDE in the unified model. Near-IR observations are therefore essential to distinguish between hot dust and a non-thermal IR excess.
  }
  % conclusions heading (optional), leave it empty if necessary 
   {}

   \maketitle
%
%-------------------------------------------------------------------

\section{Introduction}
\label{sec:intro}

% Intro TDE paragraph, emphasizing NIR

When a star ventures close enough to a supermassive black hole (SMBH) the tidal forces of the SMBH can overcome the star's self gravity, causing it to be disrupted, in what is known as a tidal disruption event (TDE). If the star is completely disrupted, approximately half of the stellar material will be accreted onto the SMBH, powering a luminous flare \citep{Rees1988,Phinney1989,Evans1989}. Originally a theoretical prediction, TDEs were first discovered by surveys in X-rays \citep{Donley2002}, and are now routinely discovered in the optical, with $>$100 TDEs now identified \citep[e.g.][]{Gezari2021,vanVelzen2021,Hammerstein2023,Yao2023}. Optically discovered TDEs display a blue thermal continuum with high ($\sim$30000 K) temperatures that persist over the TDE evolution; very broad emission lines of 5000 – 15000 km s$^{-1}$ \citep[e.g.][]{Arcavi2014,vanVelzen2021}; and a smooth power-law decline \citep[e.g.][]{Hammerstein2023}. The X-ray properties of the optically selected TDEs include ubiquitous soft X-ray spectra with temperatures $\sim$10$^{6}$~K, alongside a great deal of variation in their light curve evolution \citep[see, e.g.,][]{Saxton2020,Guolo2024}. 

% Emission mechanisms, scattering, mysteries

The observed X-ray properties of TDEs are broadly consistent with thermal emission from an accretion disk, as theoretically predicted \citep[see e.g.][]{Ulmer1999,Saxton2020}. However, the ultraviolet (UV)+optical (hereafter UVO) component at early times is not consistent with a bare accretion disk, being too luminous, and having larger inferred radii than the accretion disk \cite[see e.g.][]{Gezari2021}. The observed UVO emission has been suggested to arise from: reprocessing via enveloping or outflowing material, which can be produced via the intersecting debris streams \citep{Lu2020}, an outflow arising from super-Eddington accretion \citep{Dai2018,Thomsen2022}, or a quasi-spherical envelope \citep{Loeb1997,Guillochon2014,Metzger2022}. Alternatively the UVO emission has been suggested to arise from shocks between the intersecting debris streams themselves rather than reprocessed disk emission \citep[see e.g.][]{Piran2015,Bonnerot2021}. In the case of reprocessing, various mechanisms such as bulk scattering, absorption and compton scattering will convert the disk emission to a spectral energy distribution (SED) that peaks in the UVO regime \citep[see e.g.][for details]{Roth2016,Roth2020}. Attempts to model the resulting TDE spectra have concluded that a non-thermal spectrum is expected, with excess emission in the unobservable extreme-UV wavelengths \citep{Roth2016}, and potentially in the infrared (IR) regime \citep{Roth2020,Lu2020}.

% IR echo theory
TDEs have also been observed to produce substantial emission in the IR in excess of the expected emission associated with the accretion disk, or the hot ($\sim$20000-30000 K) blackbodies observed in UVO. The IR emission has been associated with re-radiation of the UVO emission by circumnuclear dust surrounding the TDE as an IR echo. IR echoes have been observed from supernovae over several decades \citep[e.g.][]{graham1983, dwek1983}. IR echoes arising from TDEs were predicted by \citet{Lu2016} using 1D radiative transfer models, suggesting their use for probing the nuclear dust and its distribution in the galaxies hosting TDEs. \citet{vanVelzen2016} reported mid-IR variability associated with TDEs that was consistent with expectations for an IR echo and suggested that observations of such IR echoes could be used to estimate bolometric luminosities of TDEs, with further observations of IR echoes from optical TDEs following soon afterwards \citep{Jiang2016,Dou2016}. More recently, a systematic study by \citet{Jiang2021b} found that optically discovered TDEs show dust covering factors ($L_{\rm IR}$ / $L_{\rm UVO}$) of $\sim$1\%, suggesting that the optical TDE sample favours host galaxies with small amounts of dust within their nuclear regions. Most recently, samples of energetic nuclear transients discovered via IR observations \citep{mattila2018, kool2020,Jiang2021a, reynolds2022, masterson2024,masterson2025} suggest the presence of a dust obscured population of TDEs that have remained out of reach for optical surveys. 

%\tr{Anything missing? Host galaxies of TDEs?.}

% AT 2019azh as an extremely well observed TDE
\subsection{Summary of previous observations of AT~2019azh}
% UV,opt,X-rays
In this work, we will present novel near/mid-IR observations of the TDE AT~2019azh\footnote{Recently, the transient name server (TNS) updated the prefix of all classified TDEs from ‘AT' to ‘TDE'. We will maintain the AT name in this work, to maintain consistency with previous publications.}. As a particularly nearby and recent TDE, AT~2019azh has been the subject of extensive multi-wavelength observations, which we briefly summarise here. UVO observations of AT~2019azh have been presented in a number of publications \citep{Hinkle2021a,vanVelzen2021,Liu2022,Faris2023}, and collectively provide high cadence data across the entire UVO spectrum, making this TDE one of the best observed at these wavelengths. The TDE is quite luminous at peak \citep[3rd most luminous in the sample of 17 presented in ][]{vanVelzen2021}; and exhibited an early bump in its pre-peak light curve \citep{Faris2023}; however, it is overall a fairly typical optical TDE. X-ray observations of AT~2019azh reveal a late time brightening, where the X-rays are initially detected at peak, but are faint, and then brighten at $\sim$200d before fading again \citep{Hinkle2021a}. \citet{Guolo2024} find AT~2019azh to be part of a group of late-time X-ray brightening TDEs and suggest a scenario in which the X-ray emission is present but suppressed at early time due to reprocessing of high-energy emission by optically thick material, which becomes optically thin as the accretion rate drops, causing the X-rays to escape and become observable.

% host - mention 14li here?
The host galaxy of AT~2019azh is KUG 0810+227, with a redshift of z=0.022240$\pm$0.0000071 \citep{Almeida2023}, which is a post-starburst galaxy \citep{Hinkle2021a}. These galaxies are heavily over-represented as host galaxies for TDEs \citep[see e.g.][]{Arcavi2014,French2020}. Furthermore, AT~2019azh was identified as hosting an extended emission line region (EELR) through identification of extended [O~III] emission in spatially resolved spectra \citep{French2023}. These observations could indicate the presence of a luminous active galactic nucleus (AGN) approximately 10$^{4}$ years ago, which has now declined by at least 0.8 dex based on a current AGN luminosity derived from IRAS far-IR observations. Alternatively, the EELR in the host galaxy of AT~2019azh and other TDEs may be powered by an elevated TDE rate in these galaxies \citep{wevers24,Mummery2025}. There are a number of indirect SMBH mass measurements available for the host of AT~2019azh derived from the $M_{\text{BH}}-\sigma$ relation, with two independent measurements of the velocity dispersion from medium resolution spectra yielding log$(M_{\text{BH}})=6.36\pm0.43$ \citep{Wevers2020} and log$(M_{\text{BH}})=6.44\pm0.33$ \citep{Yao2023}. 

%AT~2019azh is well observed in the radio

This paper is organised as follows. Section \ref{sec:Observations} presents our novel IR observations of AT~2019azh, including both the data reduction and photometry, as well as describing the previously published data which we will make use of in our analysis. In Sect. \ref{sec:analysis}, we analyse the IR observations through SED fitting of the photometry and IR echo modelling. Sect. \ref{sec:discussion} discusses the possible interpretations of our analysis and the implications of the evidence for a non-blackbody spectrum for the TDE emission. Finally, in Sect. \ref{sec:conclusions} we summarize our findings. 

We adopt the same cosmological parameters as used in \citet{Faris2023}, namely flat $\Lambda$CDM with H$_{0}=69.6$ km~s$^{-1}$ Mpc$^{-1}$, $\Omega_{m}=0.286$, and $\Omega_{\Lambda}=0.714$ \citep{Bennett2014}. The luminosity distance to the host of AT~2019azh is 96.6 Mpc with these parameters and the observed host redshift.

%--------------------------------------------------------------------
\section{Observations and Data Reduction}
\label{sec:Observations}

\subsection{UV and Optical}
\label{subsec:UV_opt}

We make use of the UVO data presented in \cite{Faris2023} throughout this work, which combines data presented in \citet{Hinkle2021a} and \citet{Hinkle2021b} with original observations. The data are corrected for the Milky Way extinction of $A_V=0.122$~mag as described in \citet{Faris2023}. We assume no host extinction, consistently with \citet{Hinkle2021a,Liu2022,Faris2023}, noting that if there was significant host extinction, the already UV bright TDE would become extremely luminous compared to the TDE population. All the data are host-subtracted, but we note that the 6 filters from the UltraViolet and Optical Telescope (UVOT) on the Neil Gehrels Swift Observatory have host subtractions from synthetic magnitudes obtained from SED fitting of the host galaxy rather than template imaging.

\subsection{Near-infrared}
\label{subsec:NIR}

We obtained four epochs of near-IR (NIR) imaging in the $JHK$ bands with the NOTCam instrument mounted on the Nordic Optical Telescope (NOT) as part of the NUTS2\footnote{\url{https://nuts.sn.ie/}} programme, as well as NOTCam template imaging long after the transient had faded (27 December 2023). % under similar or better conditions, to obtain seeing and depth comparable to the best quality follow-up images.
The NOTCam data were reduced using a slightly modified version of the NOTCam {\sc quicklook} v2.5 reduction package. The reduction process included flat-field correction, a distortion correction, bad pixel masking, sky subtraction and finally stacking of the dithered images. 

One epoch of $JHK$ imaging was obtained with the SOFI instrument mounted on the New Technology Telescope (NTT) as a part of the extended Public ESO Spectroscopic Survey of Transient Objects \citep[ePESSTO;][]{Smartt2015}. The reduced data was obtained from the ESO archive. Unfortunately, we did not obtain template imaging with SOFI+NTT before the instrument was decommissioned.

\subsection{Mid-infrared}
\label{subsec:MIR}

As noted in \citet{Faris2023}, AT~2019azh was observed in the mid-IR (MIR) by the Wide-field Infrared Survey Explorer (WISE) satellite  as part of the Near-Earth Object Wide-field Infrared Survey Explorer reactivation mission \citep[NEOWISE;][]{Mainzer2014}. We obtained time-resolved co-adds of the NEOWISE data of AT~2019azh \citep{Meisner2018}\footnote{\url{https://portal.nersc.gov/project/cosmo/temp/ameisner/neo8/}}. These are produced using an adaptation of the unWISE code, which stacks the individual exposures from each NEOWISE visit, without utilising the intentional blurring that is performed in the NEOWISE data, for maximum depth \citep[for more details, see][]{Lang2014,Meisner2017}.

\subsection{Photometry}
\label{subsec:photometry}

Point spread function (PSF) photometry on the NIR data was performed with the {\sc autophot} pipeline \citep{Brennan2022} after template subtraction and the resulting magnitudes were calibrated using the 2MASS catalogue \citep{Skrutskie2006}. Images were aligned for template subtraction using standard IRAF tasks and template subtraction was performed with the {\sc hotpants} package\footnote{https://github.com/acbecker/hotpants}, an implementation of the image subtraction algorithm by {\citet{Alard1998}}. Template subtracted images were visually inspected to ensure that field stars subtracted cleanly and no dipole-type residuals were left at the transient location, which would indicate a poor subtraction. Additionally, residuals at the transient location were inspected to ensure their full width at half maximum (FWHM) matched the measured seeing in the image and that they were fit well by our model PSF. For non-detections, a 3$\sigma$ limiting magnitude was obtained by injecting and recovering sources around the transient location in the template subtracted images. The NOTCam images obtained on 2019-10-23 and 2019-11-16 display a large degree of PSF elongation, likely due to issues with the telescope guiding. We were still able to obtain subtractions in which the field stars cleanly subtracted but the residual TDE PSF is elongated and so we used aperture photometry to recover magnitudes from these data. 

The $JHKs$ filters mounted in the SOFI and NOTCam instruments are not identical, which introduces systematic uncertainty into the template subtraction of the SOFI images. A comparison of the normalised filter transmission curves for the SOFI and NOTCam filters is shown in Fig. \ref{fig:filter_functions}. The $H$ and $Ks$ filters are very similar, but the $J$ filters are quite different, with the SOFI filter being considerably wider. The atmospheric transmission in the additional region covered by the red side of the SOFI $J$ filter is very low, but there is some transmission in the blue side. We do not expect the effective transmission to be extremely different, and make no attempt to correct the photometry. %We make no attempt to correct the photometry of this single $J$ band observation, but discuss possible effects of the systematic uncertainty in this measurement below.  %This is noticeable in subtractions, where the subtractions between the SOFI $HK$ data and the NOTCam templates produce smooth results, whereas the $J$ band data leave clear residuals. \tr{Need to decide what to do about this. Currently just presenting the results as they are.}

Using the publicly available NEOWISE catalog, \citet{Faris2023} find a clear detection in both the $W1$ and $W2$ bands very close to the $g$-band peak of the transient, which they find to be significantly in excess of the UVO BB, and a marginal detection in $W2$ only at a later epoch. For our MIR photometry, we performed image subtraction with the time-resolved unWISE co-adds of the NEOWISE data of AT~2019azh \citep{Lang2014}. As a template, we used the full depth unWISE co-add that includes all the data before the first detection of AT~2019azh, i.e., the ‘six-year full-depth unWISE coadds'\citep{Meisner2021}, which is the deepest template available with no possibility of flux associated with the transient. Template subtraction and photometry were mostly performed as described above for the NIR photometry, although the images are already astrometrically calibrated and did not require alignment. The results of the template subtraction are shown in Fig. \ref{fig:NEOWISE_subtractions}. The images were photometrically calibrated using the unWISE catalog presented in \citet{Schlafly2019}. We find consistent results with the detections from \citep{Faris2023}, as well as additional detections that are described below.

We correct the IR photometry for Milky Way extinction using the \citet{Fitzpatrick1999} extinction law with the corresponding coefficients taken from \cite{Yuan2013}, as in \citet{Faris2023}. We assume no additional host extinction, consistent with \citet{Faris2023} and \citet{Hinkle2021a}. We additionally follow \citet{Faris2023} in presenting all phases relative to their $g$-band peak brightness of MJD 58565.16$\pm$0.62. All phases are presented in the rest frame unless otherwise specified. The results of the photometry are given in Tab. \ref{tab:IR_photometry}. 

\section{Analysis} \label{sec:analysis}
\subsection{Evolution of the IR light curves}
\label{subsec:LC_evolution}

\begin{figure*}
   \centering
   \includegraphics[width=\textwidth]{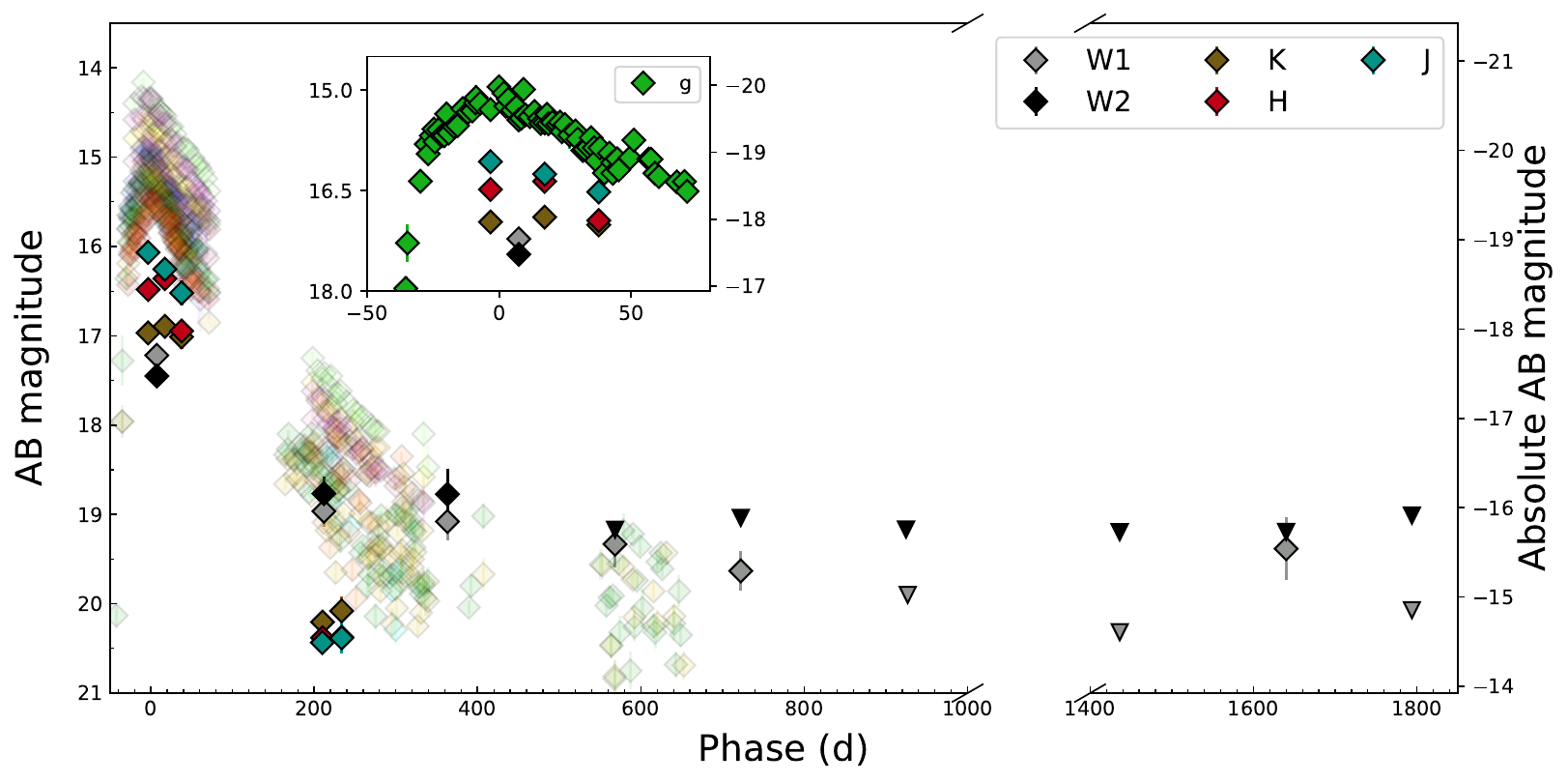}
      \caption{Photometry of AT~2019azh. IR photometry is shown with solid coloured symbols, while the extensive UVO photometry is shown using partially transparent symbols. Upper limits are indicated with downward facing triangles. The x-axis is broken, the omitted time period contains only upper limits. The inset plot shows the optical and IR light curves around peak in more detail.
      }
\label{fig:LC}
\end{figure*}

Photometry of AT~2019azh is shown in Fig. \ref{fig:LC}. Our NIR observations span the period between -3~d and 238~d, with NEOWISE MIR detections continuing until 738d, and further upper limits continuing afterwards. Additionally, there is a single $W1$ detection at 1640~d. Our coverage of the peak of the optical light curve reveals a rise and fall in both $H$ and $K$ band, while $J$ experiences only a decline. The NIR detections after the solar conjunction at $\sim$ 210~d are $\sim3$ magnitudes fainter in all bands, while the NEOWISE measurements are $\sim$1.7 mag fainter. After the solar conjunction, the $W1$ photometry shows a flat period of at least 200~d, before a slow decline until the last detection at 738~d, with multiple non-detections afterwards. The $W2$ photometry is similarly flat for 200~d, after which the source declines beyond our detection limits.
%\begin{figure}
%   \centering
%   \includegraphics[width=0.5\textwidth]{figures/IR_echo_comparisons.png}
%      \caption{
%      NEOWISE W1 absolute magnitudes for AT~2019azh, alongside the sample of IR detected TDEs in presented in \cite{Jiang2021}. 
%      }
%\label{fig:Jiang_comparison}
%\end{figure}
In \citet{Faris2023}, the authors find that the redder TDE emission peaks later compared to bluer bands, as has been found in other TDEs \citep[see e.g.][]{Holoien2020}. We fit a second order polynomial to our $H$ and $K$ band photometry and determined the implied peak times to be 11.4~d and 14.8~d for $H$ and $K$, respectively, continuing this trend. We note however, that we only have sparse data, so there is a large uncertainty on these peak times. The $J$-band data only declines, so our peak estimate is an upper limit of 17~d.

\subsection{SED fitting with dust emission}
\label{subsec:dust_fitting}

To measure the temperature of the dust that is a possible source of the IR emission, we perform fits to the UVO+IR SED of AT~2019azh at the epochs where we have IR data. For each NIR epoch, we linearly interpolate between the closest two epochs for each UVO filter to generate the full SED, estimating the uncertainties at the interpolated epochs by adding the uncertainties of the closest two epochs in quadrature. We exclude UVO filters which do not have two observations within 5 days of the IR epoch. For the NEOWISE data, we do not interpolate between the first two epochs, as there is a large change in brightness (1.7 mag) and a 6 month gap between these observations, but we interpolate between the 2nd and 3rd epoch, as there is very little change in brightness (0.1 mag). The second NEOWISE epoch is very fortunately timed within 2~d of a NIR observation, and we assume no evolution between these NEOWISE and NIR epochs.

We simultaneously fit a linear combination of a ‘hot' blackbody which we assume describes the contribution from the direct TDE emission, and a ‘cool' blackbody which describes the contribution of the emission arising from the heated dust to the observed SED. For the ‘cool' blackbody, we use a modified blackbody of the form $B'_{\nu}(T_{\text{dust}})=\kappa_{\text{abs},\nu}B_{\nu}(T_{\text{dust}})$, where $\kappa_{\text{abs},\nu }$ is the mass absorption coefficient for the dust at frequency $\nu$, which is dependent on the choice of distribution of grain radii $a$, and the grain composition. We consider both single size distributions and multiple size distributions, where the number density of grains per grain radius is proportional to $a^{-3.5}~$\citep[the MRN grain size distribution;][]{Mathis1977} from minimum size $a=0.005$ $\mu$m up to a selected maximum size. For the modified blackbody, the observed flux density can be expressed as $F_{\nu,\text{dust}} = (\kappa_{\text{abs},\nu } M_{d} B_{\nu}(T_{\text{dust}} ))/ D^{2}$ where $M_{d}$ is the total dust mass and $D$ is the luminosity distance to the TDE. %In the case where all dust grains are of the same size \citep[see e.g.][]{Uno2023}, so we can derive $M_{d}$ from our fits for a choice of grain size and distribution.

The temperature of the cool blackbody should be less than the dust sublimation temperature, $T_{\text{sub}}$, which is dependent on the grain size, composition and number density. Graphite grains able to survive temperatures $300-500$~K larger than those of silicate and larger grains have higher $T_{\text{sub}}$ \citep{Guhathakurta1989,Baskin2018}. Dust survives at higher temperatures as the number density of grains, or equivalently the gas number density, increases, with graphite grains having $T_{\text{sub}}=1500$ and $2000$~K for gas number densities $n\sim 10^{5}$ and $10^{11}$~cm~$^{-3}$, respectively \citep{Baskin2018}. Making no assumption for the gas density, we conservatively limit the cool blackbody temperature to $<2200$~K in our fits and initially make use of the $\kappa_{\text{abs},\nu }$ for a single size distribution of 0.1 $\mu \text{m}$ graphite dust, which has a high sublimation temperature. For the fitting, we employ the {\sc emcee} python implementation of the Markov Chain Monte Carlo (MCMC) method \citep{ForemanMackey2013} to fit the points and derive uncertainties. We exclude some measurements from the fits, these are indicated in the figures. This is because the $ori$ bands have a significant excess compared to the hot blackbody implied by the UV data. The origin of the excess could be strong H$\alpha$ emission, as is indeed clear in the host and continuum subtracted spectra presented in \citet{Faris2023}. Alternatively, there could be  an intrinsic deviation from a thermal spectrum, as we will discuss below.

\begin{figure*}
   \centering
   \includegraphics[width=\textwidth]{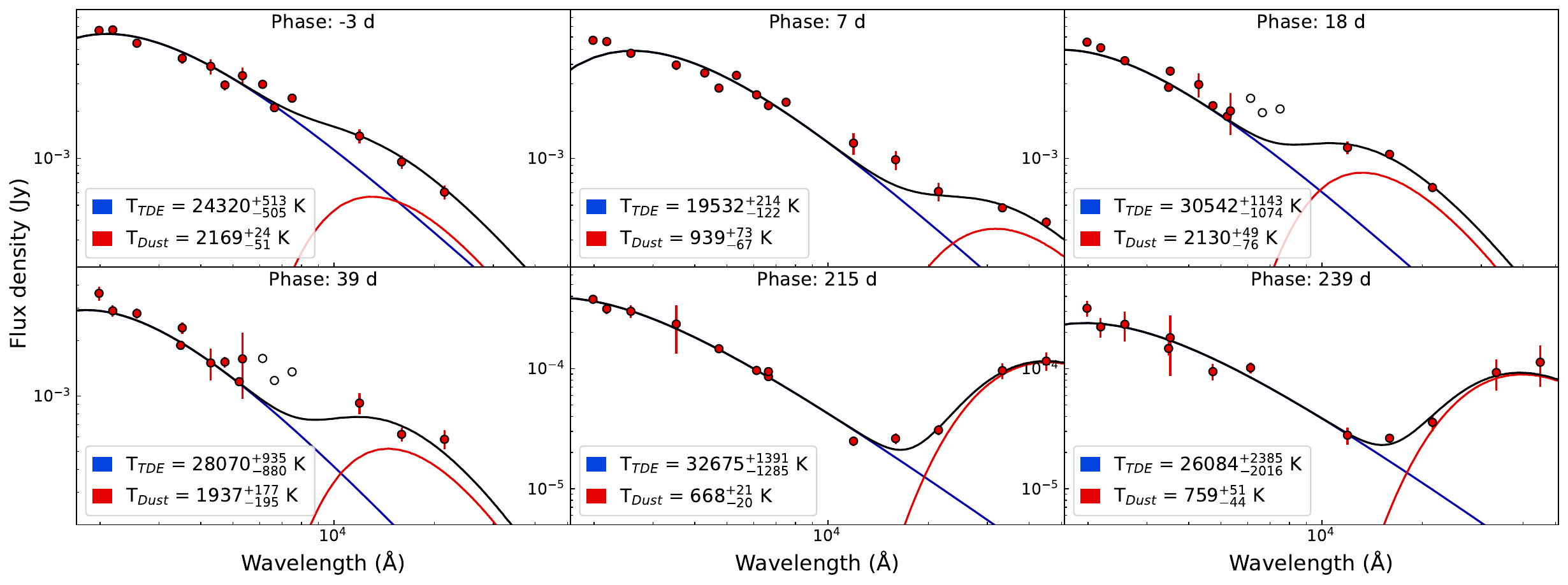}

      \caption{SED fitting with two blackbodies (hotter TDE and cooler dust) for the observed UVO+IR SEDs at 6 epochs (labeled). Fitting was performed with a modified blackbody for the dust. The dust here is entirely composed of graphite with a grain size of 0.1~$\mu$m. Empty points are excluded from the fit.
      }
\label{fig:BB_fitting}
\end{figure*}

\begin{figure}
   \centering
   \includegraphics[width=0.5\textwidth]{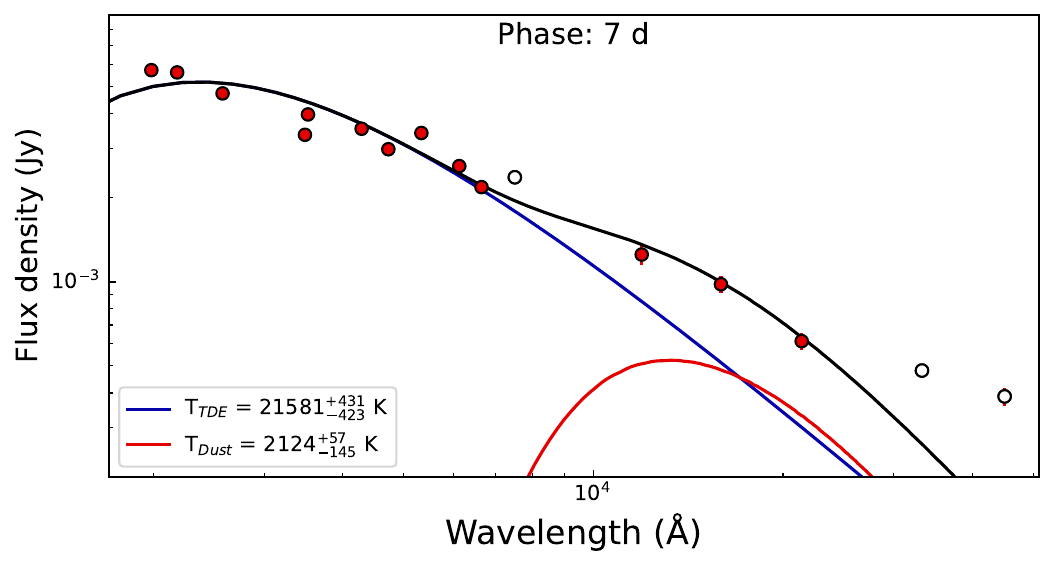}

      \caption{SED fitting with two blackbodies (hot TDE and cooler dust for the observed UVO+IR SED at 7~d. Fitting was performed with a modified blackbody for the dust. The 3.4~$\mu$m and 4.6$~\mu$m data are excluded from the fit, so that a fit with a high dust temperature is preferred.
      }
\label{fig:BB_fitting_7d_no_MIR}
\end{figure}

We show the results of these fits in Fig. \ref{fig:BB_fitting}. In all epochs, there is an IR excess compared to the hot blackbody, so the IR emission does not simply represent the Rayleigh-Jeans tail of the TDE blackbody emission. Before 40~d, we can fit the SEDs for the epochs with only $JHK$ observations using a cool blackbody with $0.1~\mu $m graphite dust. The dust temperatures are very high, but potentially consistent with dust sublimation if the gas density was sufficiently high, approximately $n\sim 10^{11}$ to $10^{12}$~cm~$^{-3}$ \citep{Baskin2018}. However, the best fit dust temperature in the second epoch at 7~d, where the SED extends to 4.6 $\mu \text{m}$, is $\sim$1000~K, much cooler than the surrounding epochs. If we exclude the measurements at 3.4 and 4.6$\mu$m and only fit the interpolated $JHK$ data with a hot dust component that is consistent with the surrounding epochs, we find a large deficit in flux density at 3.4 $\mu \text{m}$ and 4.6 $\mu \text{m}$ (see Fig. \ref{fig:BB_fitting_7d_no_MIR}). To investigate if any other dust grain size distribution can produce a good fit, we fit the second epoch with a range of $\kappa_{\text{abs},\nu }$ corresponding to graphite dust with different grain sizes, including both single and multiple size distributions. We use only graphite grains because the results of our fitting of the other epochs imply that high temperature dust is required. These fits are shown in Fig. \ref{fig:SED_fitting_many_comps}, with the reduced $\chi^2$ statistic listed. No composition yields a good fit. Although we can obtain good fits for hot $0.1~\mu$m graphite dust in epochs -3~d, 17~d, and 38~d, the additional MIR data available at the second epoch is inconsistent with dust of this temperature and composition and so we conclude that the first four epochs can not be consistently fit. 

The SEDs in the final two epochs, where the phase is more than 200~d post-peak, have a much different shape, with a clear double peak. These can be fit reasonably well with a modified blackbody, yielding cool dust temperatures of 600 - 700~K. At these cooler temperatures, a graphite dust model is not necessarily required, and using different dust grain compositions can produce a range of temperatures and masses for the dust. This will be explored further in Sect. \ref{subsec:models}. In summary, the observed SEDs of AT~2019azh are consistent with a combination of direct TDE blackbody emission and cool dust at epochs later than 200~d, but inconsistent at epochs earlier than 40~d.

\subsection{SED fitting with reprocessing models}
\label{subsec:reprocessing_SED_fitting}

% Comment out this for now. Maybe completely removed, otherwise could be in appendix.
%\begin{figure}
%   \centering
%   \includegraphics[width=0.49\textwidth]{figures/powerlawfit_Lnu.png}\newline
%   \includegraphics[width=0.49\textwidth]{figures/linear_decline_of_power_law_exponent.png}

%      \caption{\textbf{Upper panel:} Power law (L$_{\nu} \propto \nu^{\alpha}$) fits to the IR data. The red dashed lines represent power law fits to the $JHK$ bands only, while the blue dashed line represents a fit to all the IR data. The best fit power law index ($\alpha$) is listed on the plot.
      %\textbf{Lower panel:} Power law indices derived from fitting the IR SED. A linear fit to the data derived only from the IR is shown.
%      }
%\label{fig:power_law_fits}
%\end{figure}

Given the results from Sect. \ref{subsec:dust_fitting}, an alternative explanation for the observed IR emission is direct emission from the TDE, and that is not well described by a blackbody in the IR. This possibility has been discussed by a number of authors, who argue that the dominant absorption opacity at $>1$ $\mu$m can be can be due to free-free processes and produce a power-law slope that is shallower than a blackbody \citep{Lu2020,Roth2016,Roth2020}\footnote{This is analogous to continuum formation in the wind atmospheres of Wolf-Rayet stars \citep{Wright1975}}. A detailed derivation of the expected continuum spectrum is given in \citet{Roth2020}, who find that, under the assumptions of spherical symmetry and that the density drops as $r^{-n}$ with $n>1$ near the surface of the emitting material, $L_{\nu} \propto \nu^{(6-4n)/(2-3n)}$. This connects the SED shape in the IR with the density of the material from the disrupted star.

We fit our IR data at the four epochs before 40~d with a power law function $L_{\nu} \propto \nu^{\alpha}$. The measured parameters are listed in Tab. \ref{tab:power_law_fitting}, and the fits are shown in Fig. \ref{fig:power_law_fit}. For a blackbody at the observed TDE temperatures we expect the emission to follow the Rayleigh–Jeans law, $L_{\nu}\propto \nu^{2}$ at low frequencies. We observe shallower slopes at all epochs, which is visible as an IR excess in Fig. \ref{fig:BB_fitting}. The $JHK$ SED is fit well by a power law at -3 and 38~d, and fit less well at 7~d and 17~d. We note that the $JHK$ data in the 7~d fit is derived from interpolation, so this is not an independent measurement. The measured power law index $\alpha$ decreases steadily and linearly over the 40~d period of our observations, becoming shallower at a rate of $-0.015$ per day. The corresponding values of power law index for the density profile are initially very large, implying an extremely steep density profile. The power law at 7~d does not fit both the NIR and MIR part of the IR SED, with the MIR data having a different slope. This could be due to the interpolation used to estimate the NIR photometry at this epoch, evidence for excess IR emission due to dust at this epoch, or an indication that a single power law slope does not hold for our data. 

%The fit to the SED at the 7~d epoch where we also have MIR observations yields a much lower value of $\alpha$ compared to fitting the NIR alone. This is due to a break in the slope of the SED between the $K$ and $W1$ bands. Extrapolating the power law slope inferred from the NIR to the MIR and subtracting the extrapolated fluxes from the observations yields residual fluxes in W1 and W2 that are similar in brightness to the late time MIR observations, although slighter brighter (W1=18.9$ \pm $0.2 (measurement) $\pm$ 0.9 (fit); W2=18.9$ \pm $0.2 (measurement) $\pm$ 0.7 (fit) mag). It is therefore possible that there is already some contribution from dust emission at this epoch, which we consider further in Sect. \ref{subsec:models}, but the large confidence interval associated with the power law fit introduces a very large uncertainty in this estimate.

% This may go to the appendix, along with the plot
\begin{table}
\caption{Results from fitting a power law $L_{\nu} \propto \nu^{\alpha}$ to the IR SEDs observed for AT~2019azh. $\alpha_{\text{NIR}}$ was derived from fits to the $JHK$ bands, while $\alpha_{\text{IR}}$ uses the $JHKW1W2$ bands. The density profile index $n$ associated with the measurement of $\alpha$ is also listed. }
\centering
%\begin{adjustbox}{width=0.5\textwidth}
\begin{tabular}{c c c c c} \hline\hline % c|c|
Phase &  -3.04 & 7.04 & 17.81 & 38.72  \\ \hline
$\alpha_{\text{NIR}}$ & 1.42$\pm$0.06 & 1.24$\pm$0.09 & 1.1$\pm$0.35 & 0.77$\pm$0.09 \\
$n_{\text{NIR}}$ & - & 12.6 $\pm$ 11.5 & 5 $\pm$ 7  & 2.64 $\pm$ 0.32 \\ \hline
$\alpha_{\text{IR}}$ & - & 0.88$\pm$0.09 & - & - \\
$n_{\text{IR}}$ & - & 3.1$\pm$0.6 & - & - \\

\end{tabular}
\label{tab:power_law_fitting}
\end{table}

\subsubsection{The viewing angle dependent reprocessing model}
\label{subsubsec:thomsen_SED}

To further explore the non-blackbody SED of AT~2019azh, we make use of the reprocessing framework presented in \citet{Thomsen2022} and \citet{Dai2018}. This model utilises the radiative transfer code \textsc{sedona} and the General Relativistic Radiation Magneto-Hydrodynamics (GRRMHD) code \textsc{harmrad} to calculate the continuum emission from a super-Eddington accretion flow. Following the methods outlined in \citet{Thomsen2022}, we obtain a 1D density, temperature and velocity profile for a given inclination angle of the 3D accretion flow structure. This 1D structure is then converted to a spherical structure in \textsc{sedona}. The gas composition includes H, He, C, N, O, Na, Mg, Si, S, Ca, Ti, and Fe, with solar metallicity.

We initialise the radiative transfer simulation by launching millions of photon packages drawn from a blackbody spectrum, characterised by a temperature $T_0=10^6K$ and luminosity $L_{\rm inj}$. These photon packages propagate through the 3D spherical geometry and undergo scattering, and absorption and re-emission events, transforming the initial blackbody to a reprocessed spectrum. SEDONA incorporates and models various emission and absorption processes including Compton scattering, free-free, bound-free, and bound-bound transitions \citep{Kasen2006,Roth2016}. Due to the significant energy contained in the radiation, all calculations are considered under non-local thermal conditions when calculating the ionization and excitation. SEDONA is run iteratively at least 20 times to ensure the temperature has converged.

\begin{figure}
   \centering
   \includegraphics[width=0.49\textwidth]{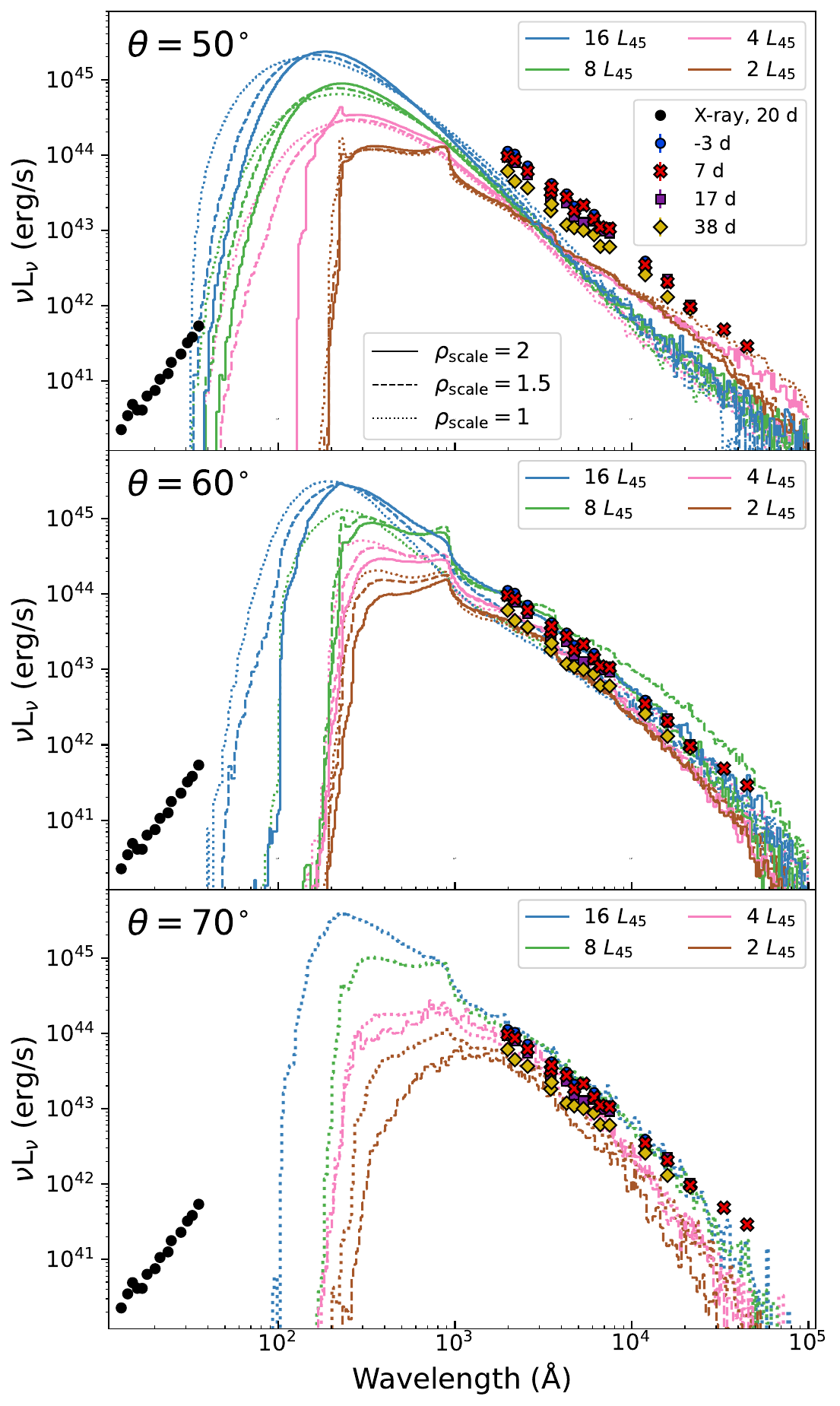}\newline
      \caption{Observed SEDs of AT 2019azh at epochs before 40~d, compared to the grid of simulated escaping spectra. The models have a wavelength grid such that the wavelength changes by 2\% at each step, and are smoothed with a median filter of size 3. The top, middle, and bottom panels show models with 50\textdegree, 60\textdegree, and 70\textdegree~ inclinations, respectively. The colouring denotes the value of $L_{\rm inj}$ as described in the legend, while the solid, dashed, and dotted lines represent models with a density profile scaling factor of 2, 1.5 and 1, respectively. We additionally show the X-ray spectrum observed at 20~d \citep{Guolo2024} with black circles.
     }
\label{fig:all_models}
\end{figure}

We produce a grid of models with viewing angles $\theta \in [50^{\circ},60^{\circ},70^{\circ}]$; luminosities $L_{\text{inj}} \in [2,4,8,16]~L_{45}$, where $L_{45}=10^{45}$ erg s$^{-1}$ and accretion rates \macc~$\in [7,12,24]$~\medd.  As we lack GRRMHD models with higher values of \macc~than 24, we multiplicatively scaled the density profile of the \macc~= 24~\medd~case to approximate higher mass accretion rates, and produce models with density scaling parameters $\rho_{\text{scale}} \in [1,1.5,2]$. The grid is incomplete for 70\textdegree~, where the extremely high densities for $\rho \in [1.5, 2]$ caused the simulations to fail in most cases.  In the GRRMHD simulations used to create the super-Eddington disk, the SMBH has $M_{\text{BH}}$ = 10$^{6}{\rm M}_{\odot}$ and spin parameter $a=0.8$. % and, as discussed in \citet{Thomsen2022}, a mass accretion rate of 24~\medd~corresponds to 46~d after the peak of the TDE for a solar-type star fully disrupted by such a black hole. 
We found that models with \macc~$\in [7,12]$ were not luminous enough to fit the data. This is consistent with the physically expected mass accretion rates - we discuss this below. 

\begin{figure}
   \centering
   \includegraphics[width=0.49\textwidth]{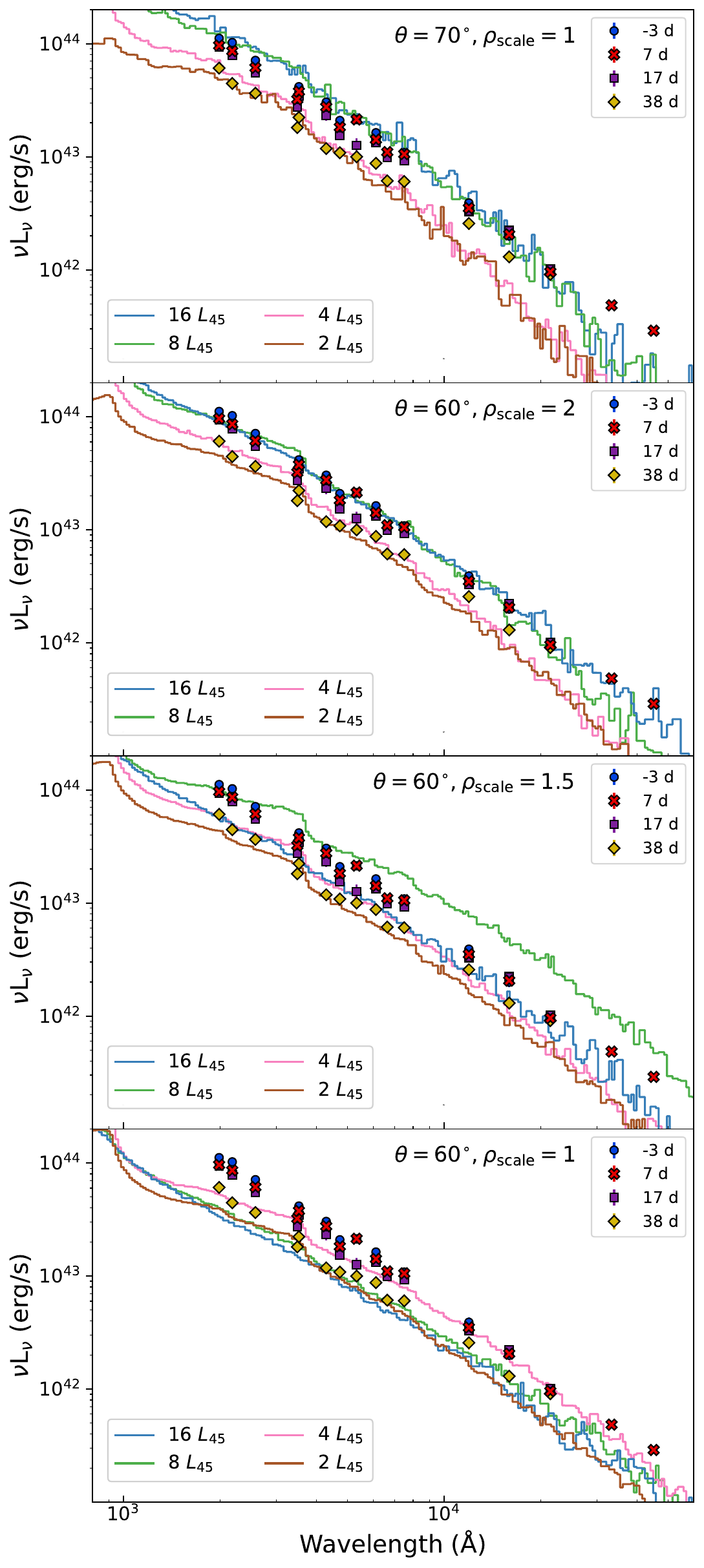}
      \caption{As in Fig. \ref{fig:all_models} but zoomed into the UVO+IR region. The top panel shows the 70\textdegree~inclination, whereas the other three panels have an inclination of 60\textdegree. The density profile scaling is indicated on the panels. 
     }
\label{fig:models_zoomed}
\end{figure}

Figure \ref{fig:all_models} shows all the models with \macc~$=24$ produced in our grid, alongside the observed data of AT~2019azh. The models reproduce the trends discussed in \citet{Thomsen2022}. For a fixed value of $L_{\rm inj}$, the spectra emit more in the UVO as the density of the reprocessing envelope along the line of sight increases, which occurs either through higher accretion rates (i.e. increasing $\rho_{\rm scale}$), or increasing the observed viewing angle. The 50\textdegree~inclination angle is not able to reproduce the observed luminosity of AT~2019azh at any epoch. For smaller inclination angles, the spectrum becomes more X-ray dominated, so we would expect even lower luminosities in the UVO+IR region. We therefore rule out viewing angles of 50\textdegree~or less. For 60\textdegree~and 70\textdegree, the models are able to reproduce the observed luminosity in the UVO region. 

In Fig. \ref{fig:models_zoomed} we zoom in on the UVO+IR part of the spectrum for the 60\textdegree~and 70\textdegree~models. The NIR slope for AT~2019azh is shallower than the Rayleigh-Jeans part of a blackbody spectrum, and to reproduce this, we require a strong contribution from free-free emission. The strength of the free-free opacity is closely tied to the ionisation state, which decreases with increasing density (i.e. $\rho_{\rm scale}$ or $\theta$), and increases with $L_{\rm inj}$. We discuss this in detail in Appendix section \ref{appendix_sec:free-free}, and show the contribution of the free-free opacity and emission in Fig. \ref{fig:free-free_on-off}. In summary, for a significant contribution from free-free emission, the density and injected luminosity must be such that the ionisation level is high enough for bound-free processes not to dominate, but not so high that electron scattering dominates. In these latter two cases, the spectrum becomes more blackbody-like. 

For an inclination angle of 70\textdegree, the UVO data around peak is fit reasonably well with $L_{\rm inj}=16$ or $8~L_{45}$, although the model spectrum is slightly too luminous. For the 38~d epoch, the UVO part of the spectrum is fit with $L_{\text{inj}}=2$ or $4~L_{45}$. In both cases, the IR data are too faint, with the MIR emission not fit well at peak, and the NIR data not fit well at 38~d. This is because the ionisation level is not high enough to produce the free-free emission required to reproduce the IR spectrum. In contrast, the 60\textdegree~model with $\rho_{\text{scale}}$ = 2 and $L_{\text{inj}}=16~L_{45}$ is able to fit the entire spectrum at 7~d including all the IR data.

In Fig. \ref{fig:all_models} we also show the observed X-ray spectrum of AT~2019azh at 20~d, reported in \citet{Guolo2024}. All of our models are much fainter than the observed X-rays. This could be explained by the assumption of spherical symmetry inherent in our 1D models, which prevents us from capturing 2D/3D effects such as photons scattering from different viewing angles to the line of sight \citep[see][]{Parkinson2024}. We discuss this further in Sect. \ref{subsec:non_BB_emission_discussion}.

%Our model grid is sparse, so we simply
%The 70\textdegree~models  with $L_{\text{inj}}>8\times10^{45}$ erg s$^{-1}$ can reproduce the luminosity at peak relatively well in the optical and UV. The fit is best for $L_{\text{inj}}=32\times10^{45}$ erg s$^{-1}$, as the models with lower values are somewhat too luminous. However, no model well reproduces the MIR data at 7~d. Additionally, although the UV and optical SED at 38~d can be relatively well fit with the models with $L_{\text{inj}}=2/4\times10^{45}$ erg s$^{-1}$, the NIR is not bright enough.

%The 60\textdegree~inclination angle fits our data best, and the data can be reproduced with $L_{\text{inj}}$ remaining relatively constant over the 40~d period for which we have the IR SED, with some decrease in density scaling, which is a proxy for the accretion rate. All models require  \macc~$\geq$ 24.%, consistent with the fact that the epochs are all before the expected time of $\sim$46~d that \macc = 24.

%\begin{figure}
%   \centering
%   \includegraphics[width=0.49\textwidth]{figures/fig5.pdf}
%      \caption{Observed SEDs of AT 2019azh, alongside three reprocessing models. The SED is well fit at peak by the model with $\rho_{\rm scale}=2$ and $L_{\rm inj}=16~L_{45}$. The other two models assume that the density and luminosity scale linearly during the decline of the accretion rate, and fit the evolution of the TDE well.
%     }
%\label{fig:best_models}
%\end{figure}

\begin{figure}
   \centering
   \includegraphics[width=0.5\textwidth]{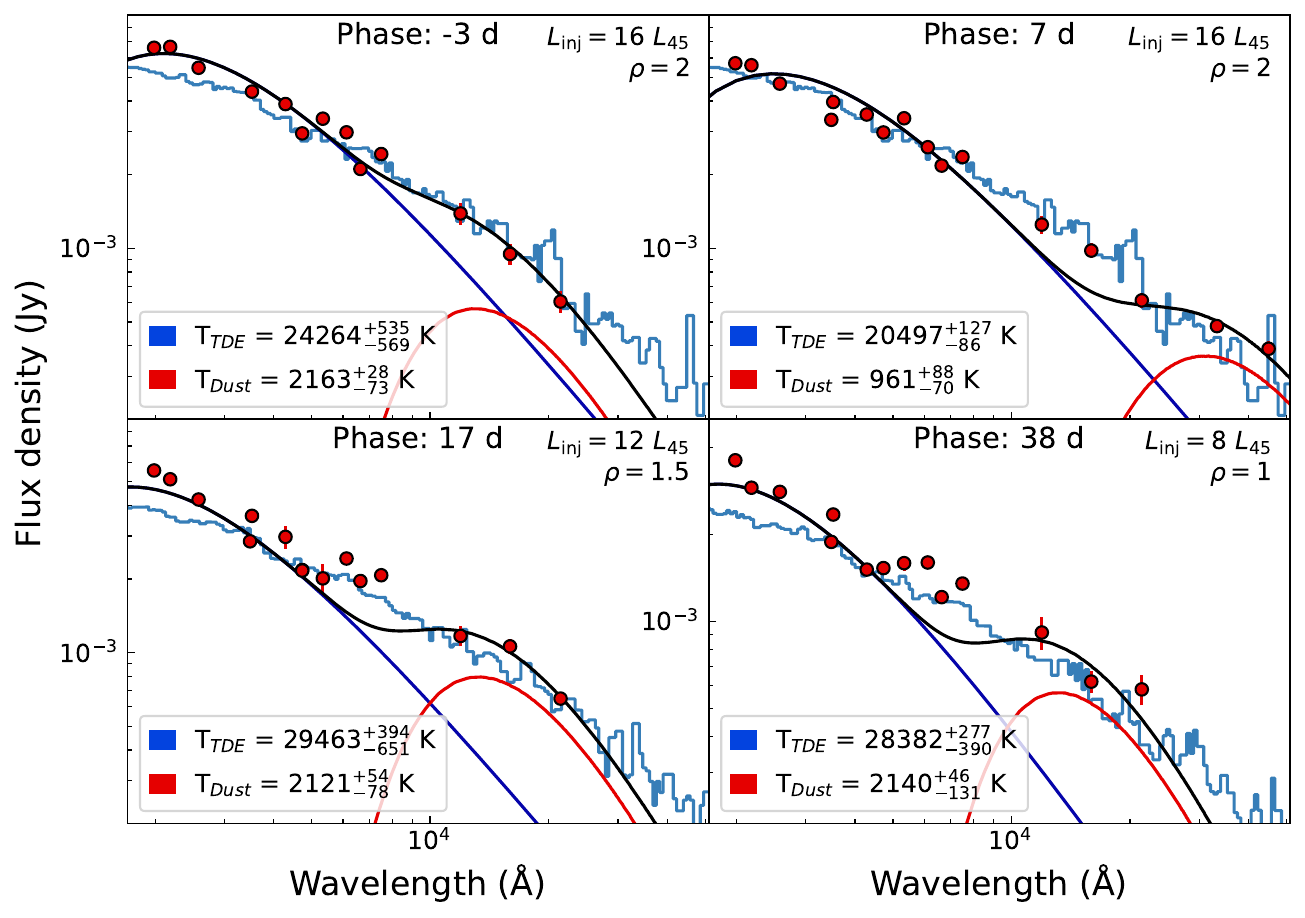}

      \caption{Observed SEDs of AT 2019azh, compared with both the best fitting dust models as in Fig. \ref{fig:BB_fitting}, and selected reprocessing models. The SED at 7~d is fit well by the model with $\rho_{\rm scale}=2$ and $L_{\rm inj}=16~L_{45}$. The models at 17 and 38~d assume that the density and luminosity scale linearly during the decline of the accretion rate. The reduced $\chi^{2}$ values associated with these models are listed in Tab. \ref{tab:model_fits}.
      }
\label{fig:fitting_comps}
\end{figure}

\subsubsection{Inferring physical parameters from spectral modelling}

From the best fitting model at peak, we can measure the accretion rate at 7~d, and use analytical expectations for the evolution of the fallback rate to select models that should fit our data at the other epochs.
%we measure the accretion rate for this TDE, and we can consider if this rate are consistent w. 
The fallback timescale for the most tightly bound disrupted material for the tidal disruption of a star with mass $m_\star$ and radius $r_\star$ is given by \citep{Evans1989,Guillochon2013,Rossi2021}: 
\begin{equation}
 t_{\rm fb}\approx 41 \ {\rm days} \ \Big( \frac{M_{\rm BH}}{10^6 {\rm M}_\odot}\Big)^{1/2} \Big( \frac{m_\star}{{\rm M}_\odot}\Big)^{-1} \Big( \frac{r_\star}{R_\odot}\Big)^{3/2},
 \end{equation}
and assuming that the fallback rate peaks at $t_{\rm fb}$, the post-peak time associated with a fallback mass accretion rate $\dot{M}_{\rm fb}$ is calculated with
\begin{equation}
\label{equation:M_fb}
\dot{M}_{\rm fb}\approx 133~ \Big( \frac{M_{BH}}{10^6 {\rm M}_\odot}\Big)^{-3/2} \Big( \frac{m_\star}{{\rm M}_\odot}\Big)^2 \Big( \frac{r_\star}{R_\odot}\Big)^{-3/2} \Big( \frac{t}{t_{\rm fb}}\Big)^{-5/3} \dot{M}_{\rm Edd}.
 \end{equation}
Some quantity of the material that falls back will be lost to outflows in the form of a disk wind, so $\dot{M}_{\rm fb} = \dot{M}_{\rm acc}+\dot{M}_{w}$. The wind value for \macc~$=24$~\medd~ measured in the hydrodynamical simulations from \citet{Thomsen2022} was $\dot{M}_{w}=14$~\medd~, implying $\dot{M}_{\rm fb} = 38$~\medd~ for the total fallback rate. To estimate $\dot{M}_{\rm fb}$ for our models with $\rho_{\rm scale} > 1$, we assume that \macc~ and $\dot{M}_{w}$ scale linearly with $\rho_{\rm scale}$, so that $\dot{M}_{\rm fb} = [57, 76]$~\medd~ for $\rho_{\rm scale}=1.5, 2$. The results from \citet{Thomsen2022} show an increase of the ratio $\dot{M}_{w}/\dot{M}_{\rm acc}$ with increasing \macc~, so our approximation for $\dot{M}_{\rm fb}$ is a lower limit.

\begin{table}
\caption{Selected models for the SEDs of AT~2019azh with IR data. The upper table lists the parameters and reduced $\chi$-squared for the dust models, while the lower table lists these values for the reprocessing models. Where two values are listed for the reduced $\chi$-squared, the former indicates the value without the $ori$ bands that were excluded from the blackbody fitting, while the latter indicates the value with all the data included.}
\centering
\begin{tabular}{c c c c } \hline\hline % c|c|
Phase & Dust temperature & Mass & Reduced-$\chi^{2}$   \\ 
(d) & K & M$_{\odot}$ &    \\ \hline
-3  & 2169$^{+24}_{-51}$ & 5.1$^{+0.5}_{-0.6}\times10^{-6}$ & 12 \\
7   & 939$^{+73}_{-67}$  & 2.1$^{+0.6}_{-0.8}\times10^{-4}$ & 14 \\
7   & 2124$^{+57}_{-145}$  & 5.2$^{+1.6}_{-0.8}\times10^{-6}$ & 26 \\
17   & 2130$^{+49}_{-76}$ &  7.9$^{+0.7}_{-1}\times10^{-6}$ & 5 / 93 \\
38  & 1937$^{+177}_{-195}$ &  8$^{+3}_{-5}\times10^{-6}$ & 3 / 60 \\
215 & 668$^{+21}_{-20}$ &  3.4$^{+0.8}_{-0.9}\times10^{-4}$ & 2 \\
239 & 759$^{+51}_{-44}$ &  1.5$^{+0.6}_{-0.9}\times10^{-4}$ & 3 \\
\end{tabular}

\begin{tabular}{c c c c c c} \hline\hline % c|c|
Phase & $L_{\rm inj}$ & $\dot{M}_{\rm acc}$ &$\rho_{\rm scale}$ & $\theta$ & Reduced-$\chi^{2}$   \\ 
(d) & $L_{45}$ & $\dot{M}_{\rm Edd}$ & & \textdegree  & \\ \hline
-3  & 16 & 24 & 2 & 60 & 25 \\
-3  & 8 & 24 & 2 & 60 & 14 \\
7   & 16 & 24 & 2 & 60 & 16 \\
17   & 12 & 24 & 1.5 & 60 & 23 \\
38  & 8 & 24 & 1 & 60 & 29 \\
% & 1000 & 1 & 100 \\
% & 1000 & 1 & 100 \\

\end{tabular}
\label{tab:model_fits}
\end{table}

Our best fitting model at -3~d and 7~d have $\rho_{\rm scale}=2$ implying that $\dot{M}_{\rm fb}(t_{\rm fb})=76$~\medd~. We adopt the SMBH mass of $M_{\rm BH}=10^{6.4}~{\rm M}_{\odot}$ derived from the velocity dispersement measurements of \citet{Wevers2020} and \citet{Yao2023}, and assume a simple stellar mass-radius relation of $R_\star/R_\odot = (M_\star/{\rm M}_\odot)^{0.8}$. % for $M_\star/M_\odot<1$ and $R_\star/R_\odot = (M_\star/M_\odot)^{0.57}$ for $M_\star/M_\odot>1$. 
We also assume that the fallback rate peak coincided with the -3~d epoch. Equation \ref{equation:M_fb} then implies a mass for the disrupted star of $M_\star\sim2.8$~M$_{\odot}$. With this value for $M_\star$, we derive expected fallback rates at 17~d and 38~d of $\dot{M}_{\rm fb}(t_{\rm fb}+17)=55$~\medd,~$\dot{M}_{\rm fb}(t_{\rm fb}+38)=40$~\medd~. These values are rather close to the expected $\dot{M}_{\rm fb}$ values of 57 and 38~\medd~for $\rho_{\rm scale}=1.5$~and 1 respectively, so we expect these models to fit these epochs well.

As we assume that the injected luminosity arises from accretion, the values for $L_{\text{inj}}$ and $\dot{M}_{\text{acc}}$ are not independent. We make the simple approximation that $L_{\rm inj}$ scales linearly with $M_{\rm acc}$ to select models that we expect to best fit the epochs at 17~d and 38~d. These models are shown in Fig. \ref{fig:fitting_comps}, and fit the evolution of the TDE SED quite well. The optical and IR data in particularly are fit well, but the UV data are somewhat more luminous than the models. Note that the model with $\rho_{\rm scale}=1.5$ and $L_{\rm inj}=12~L_{45}$ was not part of our initial grid.

The total luminosity available for injection is limited by the accretion rate, with $L_{\text{inj}} \leq L_{\text{max}} = \epsilon \dot{M}_{\text{acc}} c^{2}$ for some efficiency $\epsilon$. For $\dot{M}_{\text{acc}} = 24,36,48~ \dot{M}_{\text{Edd}}$, corresponding to assumed accretion rates for our values of $\rho_{\text{scale}}$, and with the estimated SMBH mass of $M_{\rm BH}=10^{6.4}~{\rm M}_{\odot}$ then $L_{\text{max}} = 24,36,48~L_{\text{Edd}} \sim 8,11,15~L_{45}$. Our best-fitting model at peak has $L_{\text{inj}} = 16~L_{45}$, broadly consistent but slightly larger than $L_{\text{max}}$. Reducing the value of $L_{\text{inj}}$  to be consistent with this limit would not strongly effect the resulting spectral model. %Our model grid is sparse, with no models with $8~L_{45}<L_{\rm inj}<16~L_{45}$, and the model with $L_{\rm inj}$=8$~L_{45}$ actually fits most of the data quite well, so it is likely that a denser grid would reveal equally well fitting models with lower value of $L_{\rm inj}$, resolving any tension here.
% For $\rho_{\text{scale}}=1.5$, this value is somewhat too large.
%The model with $L_{\text{inj}}=8\times10^{45}$ erg s$^{-1}$ and $\rho_{\rm scale}=1$ which well fits the 38~d epoch, is also compatible with this $L_{\text{max}}$.

To explain the mass accretion rates with a lower mass star, we would require a lower mass SMBH. For a solar mass star, the SMBH mass required for the best fit peak $\dot{M}_{\rm fb}$ is $M_{\rm BH}\sim10^{6.16}~{\rm M}_{\odot}$, which is well within the uncertainties on our SMBH mass measurements. %For these parameters, $\dot{M}_{\rm fb}(t_{\rm peak}+38)\sim30$~\medd~, which is somewhat lower than our best fit value. 
As $L_{\text{max}}\propto M_{\rm BH}$, so $L_{\text{max}}$ would also be halved for this smaller SMBH compared to our adopted value and so we would need to half the values of $L_{\rm inj}$ to maintain physical consistency. The corresponding models are not luminous enough to fit the spectrum well, particularly in the IR. However, a somewhat lower value for $M_{\rm BH}$ and corresponding lower value of $M_{\star}$ can not be ruled out. 

The luminosity we observed, i.e., the escaping luminosity $L_\text{esc}$, is less than the injected luminosity, as energy is lost to work done driving outflowing TDE material. Typical values of $L_\text{esc}$, measured directly from the model spectrum, are $\sim0.1-0.2\times L_{\text{inj}}$, depending on the specific model. For example, the best-fitting model at 7~d with $\rho_{\text{scale}}=2, L_{\text{inj}}=16~L_{45}$, and $\theta=60$\textdegree~, has $L_\text{esc} / L_{\text{inj}}=0.2$. There is a trend for this ratio to be larger for larger values of $L_{\text{inj}}$. The difference between $L_\text{esc}$ and the luminosity of the blackbody fit to the UVO data (measured in Sect. \ref{subsec:dust_fitting}), $L_\text{BB}$, represents the ‘missing energy' that we do not directly observe from the TDE. The best fitting model at 7~d has $L_\text{BB} / L_{\text{esc}}=0.05$, i.e. the missing energy is 95\% of the luminosity of the TDE at this epoch. %However, the calculation of $L_\text{BB}$ assumes spherical symmetry, whereas within the viewing angle dependent model the TDE emission is not isotropic, and therefore this estimate is an oversimplification. \tr{Could elaborate here. The point is that each viewing angle will have its own measurement.}

In order to directly compare whether the dust or reprocessing model best describes the data, we measure reduced $\chi$-squared values for the best fitting dust models and the selected reprocessing models that we show in Fig. \ref{fig:fitting_comps}. The results are shown in Tab. \ref{tab:model_fits}. The quality of fits is comparable between the two models for the first two epochs, although the hot dust model at 7~d is a worse fit. The third and fourth epochs are less well fit by the dust model if all the data, including the $ori$ bands, is included. We note that the reprocessing models come from a sparse grid, so we would expect to find better $\chi$-squared values if are grid was more densely sampled.

In summary, these reprocessing models can reproduce the entire UVO+IR SED of AT~2019azh, requiring a viewing angle of ~60$^{\circ}$; highly super-Eddington accretion, \macc~>~24~\medd; and a high mass star, with $M_{\star}>2$. The time evolution can be well reproduced under a set of simple assumptions for the fallback accretion rates. We discuss the implications of these results and their consistency with other observations of this TDE in \ref{subsec:non_BB_emission_discussion}.

\subsection{IR echo modelling}
\label{subsec:models}

The most common explanation for an IR excess observed in TDEs is an IR echo from circumnuclear dust, and many IR echoes associated with optical TDEs have been observed \citep[see e.g.][]{vanVelzen2016,Jiang2016,Jiang2021b,Dou2016}. As shown in Sect. \ref{subsec:dust_fitting} and \ref{subsec:reprocessing_SED_fitting}, the IR brightness before 40~d is not consistent with single-temperature dust, but instead non-thermal TDE emission. However, we detect an IR excess at epochs later than 200~d which is consistent with dust emission, and we can model this as an IR echo. At late times, we no longer expect a non-thermal TDE spectrum due to free-free emission, as the accretion rates and thus densities are much lower. We therefore assume that the TDE's UVO SED is well modelled with a blackbody and subtract the flux density associated with this blackbody at each epoch from the IR data to obtain the flux density associated only with the IR echo. The blackbody parameters are taken from the results of our blackbody fitting in Sect. \ref{subsec:dust_fitting}, and the resulting IR echo light curve evolution is shown in Fig. \ref{fig:IR_echo_models}. The observed $J$-band flux density after 200~d is consistent with the UVO blackbody, and is not shown. We do not propagate the uncertainties on the BB fitting.

\subsubsection{Thin shell}
\label{subsubsec:thinshell}

We first attempt to model the IR echo with a spatially-thin spherical shell of dust with radius $R$. We follow \citet{Maeda2015}, and summarise here the key points. The dust shell has a temperature $T(\theta)$, with $\theta$ the angle between the observer line-of-sight and the vector directed from the centre of the shell to the observed shell volume element, and mass $M_{d}$. Then the luminosity of the IR echo is:

\begin{equation}
\label{equation:L_theta}
    L_{\nu,\text{echo}} = 2\pi M_{d}\kappa_{\text{abs},\nu}\int_{0}^{\pi}\text{sin}~\theta~B_{\nu}(T(\theta))\text{d}\theta,
\end{equation}

\noindent where $\kappa_{\text{abs},\nu}$ is the mass absorption coefficient of the dust at frequency $\nu$, which is determined by the dust properties. We assume a uniform number density for the shell. %, and that the dust is graphite, as this is the most resistant to evaporation.
We can derive a time-dependent luminosity by relating $t'$, the time at which light emitted from a given volume element at $\theta$ is observed, to $t$, the time since the first light from the TDE, as follows:

\begin{equation}
\label{equation:delay_time}
t' = t - \frac{R}{c}(1-\text{cos} (\theta)).
\end{equation}

\noindent From Eqs. \ref{equation:L_theta} and \ref{equation:delay_time}:

\begin{equation}
\label{equation:model_time}
    L_{\nu,\text{echo}}(t) = 2\pi \frac{c}{R} M_{d}\kappa_{\text{abs},\nu}\int_{\text{max}(t-\frac{2R}{c},0)}^{t} B_{\nu}(T(t'))\text{d}t'.
\end{equation}

 %Alternatively the dust can be configured in a torus configuration, with some opening angle $\theta_{0}$. In this case, assuming that the observer is viewing the torus from the polar direction, the luminosity evolution follows:

%\begin{equation}
%\label{equation:model_time_torus}
%    L_{\nu,\text{echo}}(t) = 2\pi \frac{c}{R}\frac{M_{d}}{\text{sin}(\theta_{0})} \kappa_{\text{abs},%\nu}\int_{\text{max}(t-\frac{R}{c}(1+\text{sin}(\theta_{0})),0)}^{t-\frac{R}{c}(1-\text{sin}(\theta_{0})} B_{\nu}(T(t'))\text{d}t'
%\end{equation},

%where the integral is set to zero in the case where $t-\frac{R}{c}(1+\text{sin}(\theta_{0}) < 0$. This model produces a square wave function that rises later and then falls faster compared to the spherical shell model, with the rise coming later and falling earlier for torii with small opening angles $\theta_{0}$.
%\tr{Need some detail here about the extended dust model from Takashi, if we include it}

\noindent We additionally assume that the temperature of the dust is determined by radiative equilibrium with the TDE flux:

\begin{equation}
\label{equation:model_temp}
 \int_{0}^{\infty} \frac{L_{\text{TDE},\nu}(t)}{4 \pi R^{2}}\kappa_{\text{abs},\nu } \text{d}\nu = 4 \pi \int_{0}^{\infty}\kappa_{\text{abs},\nu } B_{\nu}(T(t))\text{d}\nu,  
\end{equation}

\noindent which determines the dust temperature $T(t)$ for a given TDE luminosity evolution and dust shell radius. As discussed in \citet{vanVelzen2016} and \citet{Lu2016}, and shown in the examples from \citet{Maeda2015}, such a model produces a square wave if the light travel time to the dust is larger than the characteristic timescale of the TDE. 

%Furthermore, the square wave function extends from $\tau = 0$ to $\tau = 2R / c$.

We assume that the pseudo-bolometric LC derived in \citet{Faris2023} is a good approximation for the total luminosity of the TDE at the wavelengths responsible for heating the dust, and use this as an input to the model. The free parameters are then the dust composition, which determines the mass absorption coefficient, and the mass and radius of the dust shell. The dust shell is spatially-thin - this is an approximation for a more physical distribution of dust, that can be extended to larger radii. This is a reasonable approximation as the IR echo signal will be dominated by the emission arising at the inner part of the distribution, where the dust is hottest, the density highest and the IR echo has a shorter duration, due to shorter light travel times.

%hen for a specific dust size distribution and composition, we can find a radius for the dust shell that best fits these equations. 

%The IR echo model peaks at $\sim$100~d, whereas the IR echo photometry peaks much earlier - before 38~d in the case of $H$ band. This peak is set by the radius of the dust shell, and can not be moved closer unless the dust has a higher sublimation temperature. The observed $JHK$ colours at peak are relatively well reproduced, but the MIR data are not well fit, being much brighter than the model predicts. Additionally, this model can not reproduce the much redder colours that we observe at late times, as the IR echo model temperature is still much higher than that implied by the observed SED. In agreement with the results of the blackbody fitting in \ref{subsec:dust_fitting}, the IR detections before 40~d are not well explained by dust emission.

\begin{figure}
   \centering
   \includegraphics[width=0.49\textwidth]{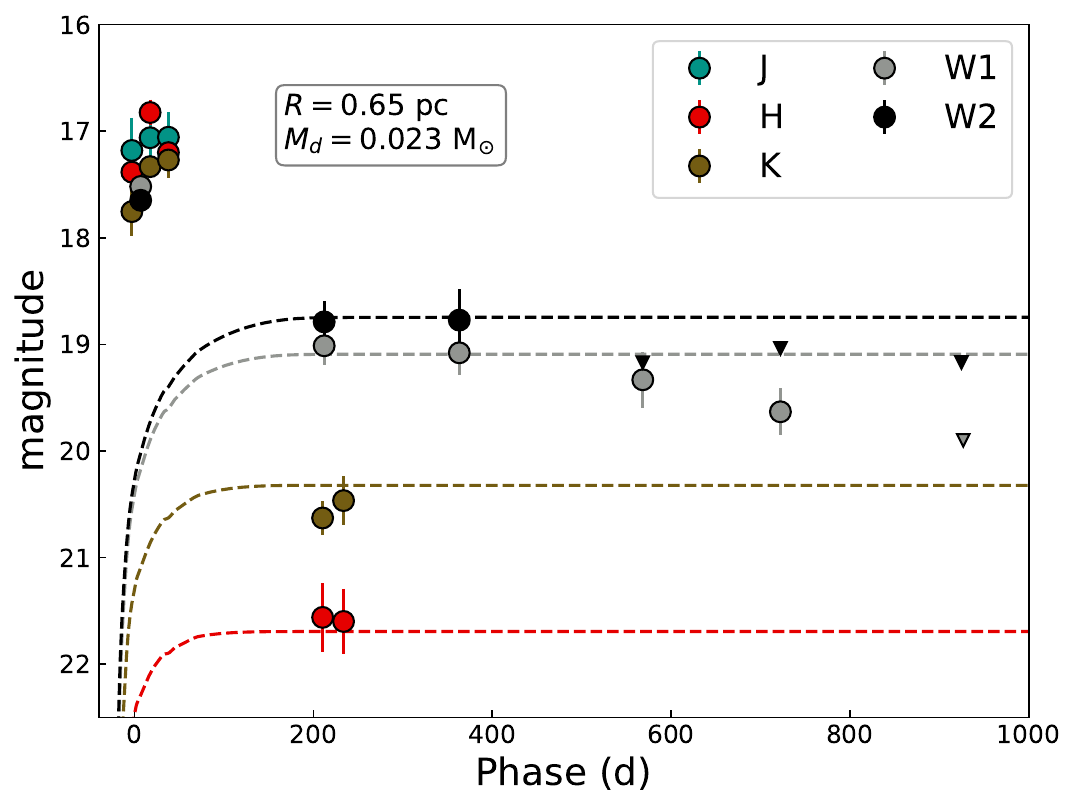} \newline
   \includegraphics[width=0.49\textwidth]{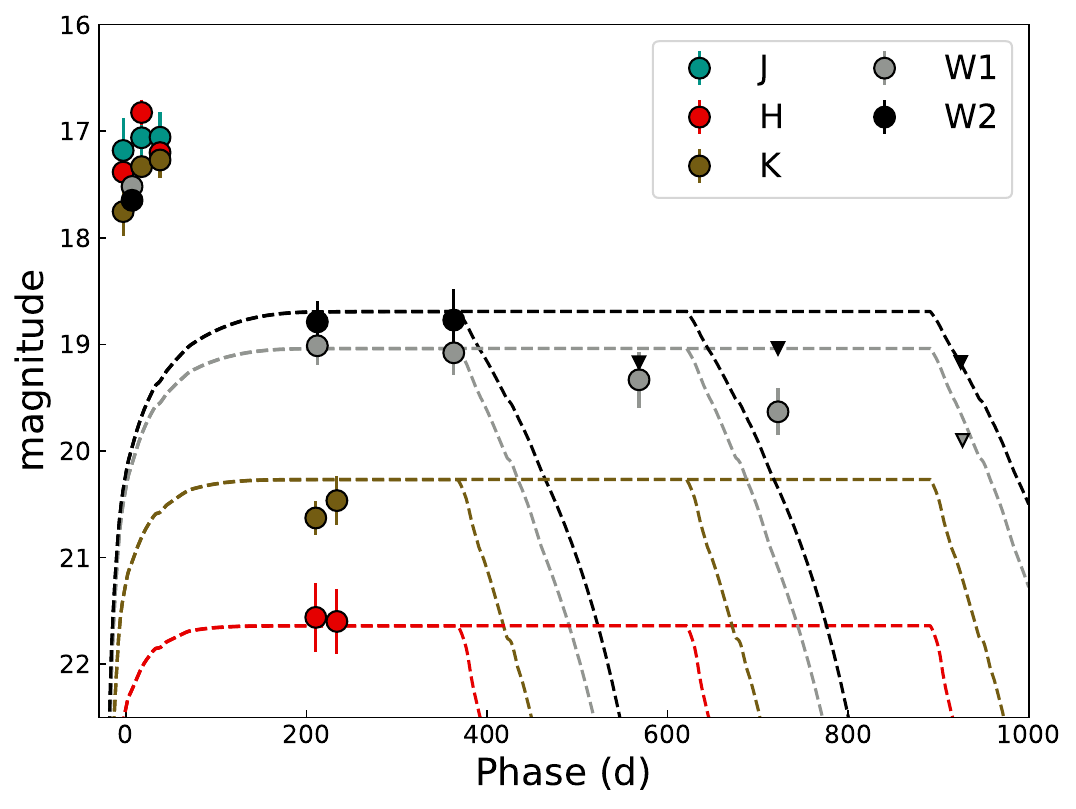} \newline
      \caption{
      \textbf{Top}: IR echo model for a spherical shell of silicate dust. The dust shell has a radius 0.65~pc, to match the observed photometry. The shell does not reproduce the observed drop in the MIR light curves. The contribution of the hot blackbody has been subtracted from the data. \textbf{Bottom}: IR echo models for different silicate dust clumps. The dust is at 0.65~pc from the TDE and has $\theta_{0}=30$\textdegree, 40\textdegree, 50\textdegree; $\theta_{\text{clump}}=60$\textdegree,~80\textdegree, 100\textdegree. The models with larger values of $\theta_{\text{clump}}$ decline at later times.
      }
\label{fig:IR_echo_models}
\end{figure}

Our SED fitting in Sect. \ref{subsec:dust_fitting} implied that the dust temperatures are low ($<1000$~K), much less than the sublimation temperature. We can not then assume that the dust composition and radius is controlled by dust sublimation as a result of the TDE. Using a graphite dust composition, we require dust at $\sim$ 1 pc to reproduce the observed colours; however, silicate dust can be located more nearby, with a good fit to the observed colours at $\sim$ 0.65 pc ($\sim$2 light years), shown in Fig. \ref{fig:IR_echo_models}. Such a large radius for the dust shell creates a long-lived IR echo (twice the light travel time, i.e. $\sim$~4 years), that does not reproduce the observed decline and drop in the MIR observations. Therefore, the spherical shell IR echo model can not self-consistently reproduce both the timescale and the colours of the IR echo we observe.

%The IR photometry after $\sim 200$~d has much redder colours and corresponding low temperatures,  implying that we are not observing dust at the sublimation temperature, and so can consider other compositions and radii for the dust.
% If we instead optimise the fit for the shape of the $W1$ observations, neglecting the colour evolution, we obtain a relatively good fit for dust at a 0.3 pc radius, shown in Fig. \ref{fig:W1_optimised_model}. However, the temperature is too hot to well fit the other bands, particularly the faint $K$ and $H$ emission. 

%In Sect. \ref{subsec:reprocessing_SED_fitting}, we fit a power law to the NIR measurements at 7~d and find that the MIR observations are in excess of this power law. To see if this excess is consistent with dust emission, we subtract the power law fit from the observed W1 and W2 fluxes, and show the resulting magnitudes in Fig. \ref{fig:IR_echo_models}. The IR echo does not rise fast enough to well fit these values, although with the very large uncertainty arising from the power-law fit, we can not say that they are inconsistent.
% blob dust model?

\subsubsection{Dust clumps}
\label{subsubsec:blob}

The temperature of the dust, and therefore the colour of the IR emission, is determined by the distance to the dust from the TDE. By altering the geometry of the dust, we can change the time evolution of the IR echo while still obtaining similar colours for the IR emission. We consider a dust clump, which we here simply approximate with a section of the overall spatially-thin spherical shell with the same properties as in Sect. \ref{subsubsec:thinshell}, at distance $r_{\text{clump}}$. The section is defined by $\theta_{\text{clump}}$, the angle between the observer line of sight and the centre of the section, and $\theta_{\text{0}}$ which is the opening angle of the section in both the polar and the azimuthal direction. The section has an infinitely-small radial thickness d$r$, the solid angle of the section is:
\begin{equation}
\begin{split}
    %\text{d}
    \Omega_{\text{clump}} = & \int r_{\text{clump}}^{2}\text{sin}\theta~ \text{d}\theta \text{d}\phi \\
    = & r_{\text{clump}}^{2} \int_{\theta_{\text{clump}}-\theta_{\text{0}}/2}^{\theta_{\text{clump}}+\theta_{\text{0}}/2}\text{sin}\theta~ \text{d}\theta  \int_{-\theta_{\text{0}}/2}^{+\theta_{\text{0}}/2}\text{d}\phi \\
    = & \theta_{\text{0}} r_{\text{clump}}^{2} [\text{cos}(\theta_{\text{clump}}-\theta_{0}/2) - \text{cos}(\theta_{\text{clump}}+\theta_{0}/2)],
\end{split}
\end{equation}
and the density of the clump is:

\begin{equation}
    \rho_{\text{clump}} = \frac{M_{d}}{ 
    %\text{d}
    \Omega_{\text{clump}} \text{d}r}.
\end{equation}

We then calculate again the luminosity of the IR echo from this clump and find:

\begin{equation}
\label{equation:clump_LC}
\begin{split}
L_{\text{echo},\nu}(t) = & \int 4 \pi\kappa_{\text{abs},\nu } B_{\nu} (T(\theta)) \rho_{\text{clump}} \Omega\, \text{d}r \\
& = \frac{4 \pi c \kappa_{\text{abs},\nu} M_d}{r_\text{clump} [\text{cos}(\theta_{\text{clump}}-\theta_{0}/2) - \text{cos}(\theta_{\text{clump}}+\theta_{0}/2)]} \\
& \quad \times \int_{\text{max}([t-r_{\text{clump}}/c(1-\text{cos}(\theta_{\text{clump}}+\theta_{0}/2))],0)}^{\text{max}([t-r_{\text{clump}}/c(1-\text{cos}(\theta_{\text{clump}}-\theta_{0}/2))],0)} B_{\nu}(T(t')\text{d}t'
\end{split}
\end{equation}

\noindent where $t'$ is defined as in Eq. \ref{equation:delay_time} and the temperature evolution by Eq. \ref{equation:model_temp}\footnote{For a more detailed derivation, see \citet{Maeda2015}.}.

To fit the IR LC after 200~d, we explore silicate dust clumps placed at approximately 0.65~pc, in order to match the observed colours of the IR echo. In Fig. \ref{fig:IR_echo_models} we show models arising from some dust clumps with $r_{\text{clump}}=0.65$~pc, $\theta_{0}=30,40,50$\textdegree~, and $\theta_{\text{clump}}=60,80,100$\textdegree~ respectively. These dust configurations produce IR echoes that rise as early as possible, to demonstrate that we expect only a small contribution from this dust at early times. The epoch at which the echo sharply declines is later for larger opening angles $\theta_{\text{clump}}$. The slow decline after $\sim400$~d is not well reproduced by a spatially thin dust shell, but would be naturally produced by extending our model to include an extended distribution for the dust \citep[see e.g.][]{Nagao2017}. Thus a dust clump with $\theta_{0}=30$\textdegree~and $\theta_{\text{clump}}=60$\textdegree, with an extended dust distribution, could likely fit these data. However, given the lack of constraints on the many free parameters of the model, we choose not to further increase the model complexity. Using Eq. \ref{equation:clump_LC}, we can calculate the dust covering factor directly by measuring the total radiated energy of the IR echo ($E_{\text{dust}}$) and comparing to the observed total radiated energy of the TDE ($E_{\text{TDE}}$). For the model with $\theta_{0}=40$\textdegree~and $\theta_{\text{clump}}=80$\textdegree, we find a covering factor $E_{\text{dust}}/E_{\text{TDE}}\sim 1$\%. The mass of dust in this model is 0.01~M$_{\odot}$.

Interestingly, there is an additional NEOWISE detection in the $W1$ band at 1640~d, shown in Fig. \ref{fig:LC}. The $W1$ mag is fainter than at 212 and 363~d, and there is no $W2$ detection. For a time lag of 1640~d, the minimum radial distance to the dust, occurring in the case that the dust is on the far side of the SMBH, is 1640/2 light days, i.e.  0.69~pc. The detection is thus consistent with the presence of an additional dust clump further out than the first, and given that the epochs immediately before and after the detection yield no detection, it should have a relatively small opening angle.

In summary, it is possible to explain the observed IR data of AT~2019azh after 200~d with an IR echo arising from a clump of silicate dust that lies $\sim~0.65$~pc from the TDE. It is important to note that these parameters are not strongly constrained. For example, using different dust compositions will change the distance to the dust (larger radius for larger proportions of graphite) and our sparse photometry does not properly constrain the rise and fall of the echo.

\section{Discussion} \label{sec:discussion}

\subsection{Reprocessing models for AT~2019azh}
\label{subsec:non_BB_emission_discussion}

%\subsection{Direct TDE emission can produce an IR excess}
%\label{subsec:non_BB_emission_discussion}

% dust doesn't work at early times
Infrared detections of TDEs have previously been interpreted as arising from an IR echo from dust. However, in the case of AT~2019azh, the IR emission at epochs before 40~d is difficult to explain as dust emission. This is primarily because we are unable to fit the SED that extends to 4.6~$\mu$m at 7~d with a blackbody corresponding to any graphite dust composition. It is possible to fit the $JHK$ SED at the other early epochs with very hot graphite dust, but a consistent temperature can not simultaneously fit the entire SED we observe at 7~d, drastically underestimating the 3.4~$\mu$m and 4.6~$\mu$m flux. While it is possible to introduce multiple dust components to better fit the SED, this would require a very fine-tuned scenario. It would require a dust component at the sublimation radius to produce the hot emission seen in the NIR, and a substantially further out dust component to produce the bright, but much cooler MIR emission. Crucially, to agree with our observations, these two dust components should be much less luminous at epochs later than 200~d, which is only possible if both lie close to the observer line of sight, as otherwise they would still provide significant emission at later times. We find such a scenario implausible.

% General predictions for IR excess in the TDE emission
The alternative explanation is that the IR emission arises primarily directly from the TDE. To produce an IR excess compared to a hot blackbody describing the TDE emission, the free-free opacity should dominate in the IR, which requires a geometrically and optically thick reprocessing envelope. The predicted spectrum follows a power law, and our NIR data can be fit well with such a power law. However, as we only have three NIR bands, this power law fit is not well-constrained. We found that the evolution of the required power law index implies a very steep density distribution outside the emitting region at peak that rapidly and linearly becomes shallower. A major powering mechanism invoked to produce the UVO emission from TDEs is stream-stream collisions. \citet{Lu2020} presents the collision-induced outflow (CIO) model for TDEs, in which self-intersection of the fallback stream unbinds a large amount of shocked gas and produces an outflow. They predict that the reprocessing in the unbound envelope will produce a shallow slope in the IR, $L_{\nu}\propto\nu^{\sim0.5}$, rather close to our observed values at $\sim39$~d.

The viewing angle dependent model of \citet{Dai2018} and \citet{Thomsen2022} can well reproduce the IR SED of AT~2019azh, while simultaneously fitting the UVO data. The best fit parameters require a relatively steep viewing angle for the observer of $\sim$60\textdegree~ and very high accretion rates of $M_{\text{acc}}>24~\dot{M}_{\rm Edd}$ at peak. We also found that to reproduce the large value of $M_{\rm fb}$ that is implied by our spectral fitting at peak, we require a high mass star to be disrupted, with $M_\star>2$M$_{\odot}$ being the preferred mass. The starburst age for the host galaxy of AT~2019azh has been measured as $200\pm30$~Myr \citep{wevers24}, and a 3~M$_{\odot}$ star has a main sequence lifetime of $\sim$300 Myr \citep{Kippenhahn2013}; therefore there is a population of such stars available to be disrupted. Light curve fitting codes have been used to estimate $M_{\star}$ for AT~2019azh: the widely used code \texttt{MOSFiT}\citep{Guillochon2018} has found very low values of 0.1 M$_{\odot}$ \citep{Hinkle2021a,Faris2023} as well as a high value of $\sim$4 M$_{\odot}$ \citep{Hammerstein2023}. Neither result is secure, with \citet{Faris2023} arguing that the low mass values are not self-consistent, and the SMBH mass estimated in the results of \citep{Hammerstein2023} being an order of magnitude larger than that derived from the velocity dispersion measurements. The high luminosity of AT~2019azh compared to the TDE population is also naturally explained if the disrupted star has a larger mass compared to the general population, and therefore a larger fallback rate. We note that there are many assumptions that go into this measurement. Some examples include: Eqn. \ref{equation:M_fb} is not expected to hold for early times when the original structure of the disrupted star still impacts the light curve evolution \citep{Guillochon2013}; the relationship between the peak optical luminosity in any particular band and the peak accretion rate is not clear; and our sparse grid means we do not have a precise measurement of the accretion rate or the inclination angle. 

%A 3This is recent enough that stars with mass up to 4-5 M$_{\odot}$ will still be evolving on the main sequence, so high mass stars are available to be disrupted.

As well as fitting the UVO+IR SED, the X-ray behaviour of AT~2019azh is also well described by the viewing angle dependent reprocessing model, with the viewing angle we obtain. The X-ray emission in this model is prompt, but the X-rays are reprocessed in the optically thick material. The optical depth through the observer line of sight depends on the viewing angle and on the mass accretion rate of material. At low viewing angles (close to the pole), X-rays efficiently escape at all times, and the X-ray luminosity is always large compared to the UVO luminosity. At high viewing angles, the opposite is the case. At intermediate angles, the mass accretion rate is important: X-rays are suppressed at early times while the accretion rates are high, but as the accretion rate drops the optical depth decreases and the ratio of the X-ray to UVO luminosity increases. X-ray observations of AT~2019azh show this behaviour, described in \citet{Guolo2024} as a late-time brightening. AT~2019azh is detected at early times (20~d) but the ratio of UVO to X-ray luminosity is of order 10$^{3}$. By 225~d, this ratio has become of order unity, and it remains so until 404~d. As previously mentioned, the 1-D nature of our modelling prevents direct comparison of the observed X-rays to our model spectra. However, the observed X-ray behaviour is consistent with the expectations for intermediate viewing angles, such as our predicted viewing angle of 60\textdegree.  The CIO model also predicts that an X-ray brightening can occur in some TDEs, due to the CIO blocking the line of sight to the accretion disk at early times. Therefore, their scenario is also qualitatively consistent with the X-ray observations of AT~2019azh. 

In \citet{Goodwin2022}, analysis of the radio observations of AT~2019azh revealed an outflow occurring at early times (10~d pre-peak), around the time of the stellar disruption. They argue that the CIO model would produce an outflow with similar properties to that inferred from the radio data, and that the early timing can be explained by the outflow being launched through the stream-stream collision. An accretion driven outflow is disfavoured by the lack of X-rays at early times unless there is significant obscuration of the X-rays, as in the viewing angle dependent model.

In summary, the multi-wavelength observations of AT~2019azh seem consistent with either the viewing angle dependent model or the CIO model for TDE emission. We have explored the viewing angle dependent model in greater depth, and show that we can reproduce the UVO+IR SED, and constrain the TDE viewing angle. We encourage further exploration of both models.

%The deviation from a blackbody due to free-free opacity in the IR is also directly predicted to be observable in the case of the collision-induced outflow model proposed by \citet{Lu2020}, and they predict a power law slope. They predict lower values for $\alpha$ than we observe in our fits to the NIR spectrum at early times, but at 38~d the measured $\alpha$ is quite consistent with their predictions. 

\subsection{Dust properties}
\label{subsec:early_time_IR_echo_discussion}

The IR emission from AT~2019azh later than 200~d is consistent with an IR echo from $\sim 0.01$ M$_{\odot}$ of dust confined to a relatively small region approximately 0.65~pc from the central SMBH. Assuming a typical gas to dust mass ratio of 100, this quantity of dust would correspond to $\sim1~{\rm M}_{\odot}$ of gas. The inner parsec around the central SMBH thus appears to be quite sparse, consistent with the nature of the host galaxy as quiescent. We additionally observe a short lived-late time IR flare that is consistent with another dust clump, at similar or greater distance than the first one. The observations are consistent with a distribution of clumpy clouds of gas and dust, similar to those observed in the Milky Way, where tens of gas clouds have been observed within 0.5-2~pc of the SMBH Sgr A*, although these clouds are much more massive ($\sim 10^4~{\rm M}_{\odot}$) \citep{Mezger1996,Christopher2005}.

Given that we have concluded that the early time IR excess is not due to dust, we never observe very hot dust in connection with AT~2019azh. This implies that the TDE itself does not sublimate a significant amount of dust, and therefore the dust distribution is independent of the TDE flare. The dust properties we infer are dependent on a number of simplifying assumptions about the TDE emission, which are not necessarily valid. First, the inferred luminosity of the TDE we use arises from blackbody fitting of the UVO observations. As discussed above, AT~2019azh was observed to produce significant X-ray emission and furthermore, it is expected that TDEs produce significant emission in the extreme-UV which we can not directly observe. Integrating the SEDs of the models we present in Sect. \ref{subsubsec:thomsen_SED}, we find that the total luminosity of the best fitting models is $\sim$10 times that of the blackbody luminosity at the same epoch. This energy can be efficiently absorbed by the dust, and will require the dust to be more distant for our IR echo model to match the observed IR SED. If the dust is much more distant, then a smaller opening angle for the dust is required to produce the same timescale flare than for closer dust.

Independent measurements of the density and location of the circumnuclear material around the SMBH hosting AT~2019azh arise from radio observations. \citet{Goodwin2022} found fluctuations in the synchrotron energy index observed for AT~2019azh, and argue that an inhomogenous circumnuclear medium could explain these observations. Additionally, \citet{Zhuang2025} argue that the multiple radio flares AT~2019azh exhibited significantly after the peak optical emission ($\sim$350 and 600~d) were produced by the collision of the TDE outflow with multiple circumnuclear gas clouds, each producing a luminous radio flare. By modelling the flare associated with the scenario, they find that these clouds should lie at 0.1~pc and 0.2~pc, and have radii of 0.04 and 0.08~pc, respectively. These clouds are much closer than we infer for the dust cloud that produces the IR echo, and are close to the lower limit for the sublimation radius for AT~2019azh. We thus would expect to see high temperatures for any surviving dust in these clouds, which we do not observe. Alternatively, if the intrinsic bolometric luminosity is higher than that inferred from the blackbody luminosity, all the dust within these clouds could be destroyed. The sublimation radius approximately increases $\propto L_{\text{max}}^{1/2}$, implying that if we are truly underestimating the peak luminosity by a factor of 10, the sublimation radius would be $\sim$0.28~pc for the graphite dust with a multiple size distribution, implying that any dust in the above clouds may be destroyed. There is thus evidence from both radio and IR for a clumpy, inhomogeneous medium, at different physical scales traced by the TDE light, and the much slower TDE outflow.

\subsection{Implications for the TDE population} 

\begin{figure}
   \centering
   \includegraphics[width=0.49\textwidth]{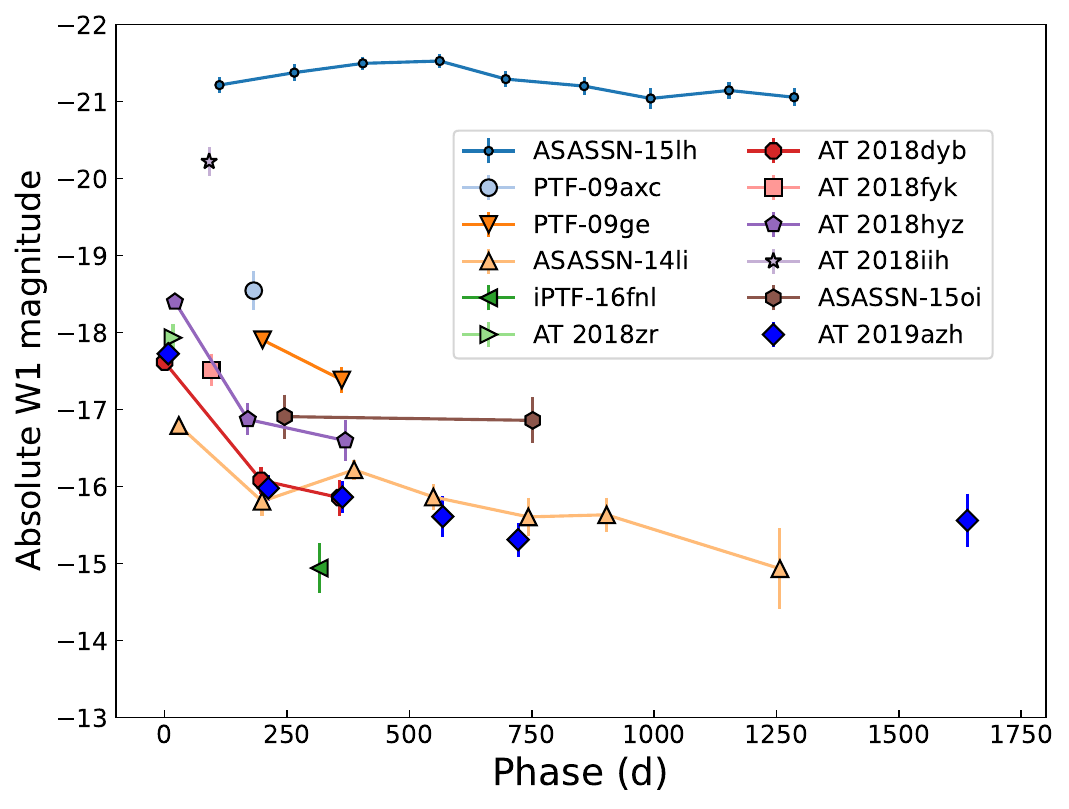}\\
   \includegraphics[width=0.49\textwidth]{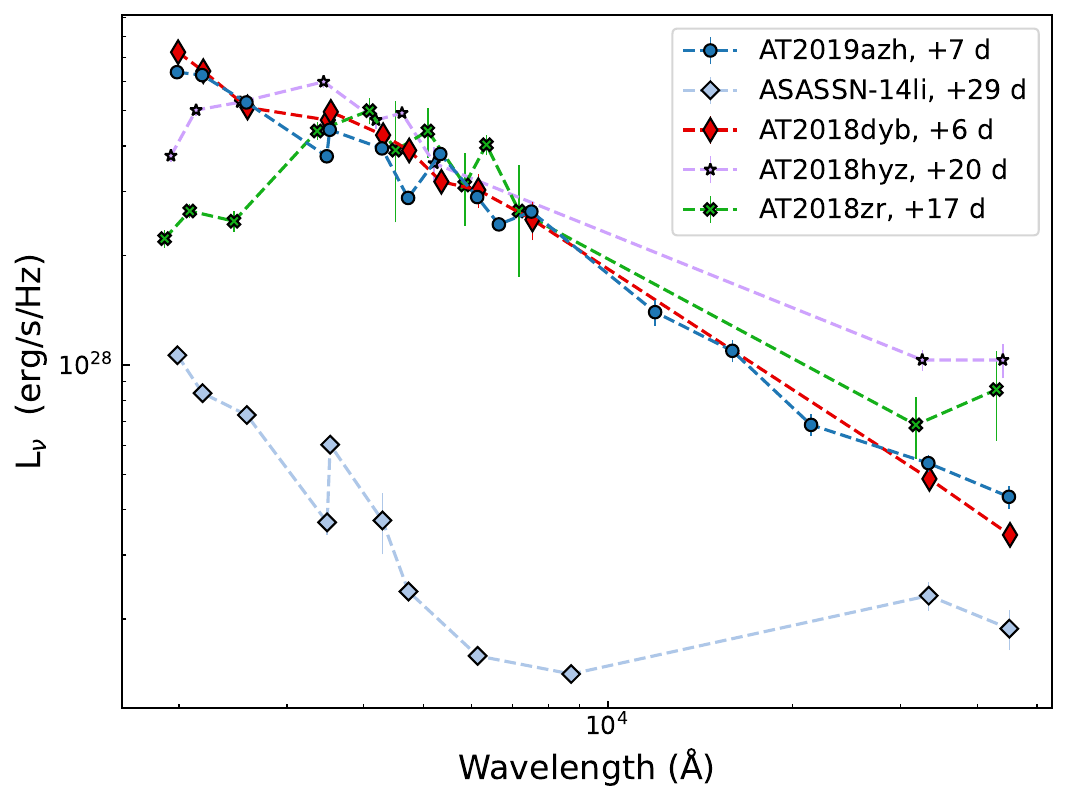} 
      \caption{Upper panel: NEOWISE W1 absolute magnitudes for AT~2019azh, alongside the sample of IR detected TDEs presented in \cite{Jiang2021b}. Lower panel: SEDs of TDEs in the literature with IR observations near to the TDEs optical peak. MIR data taken from \citet{Jiang2021b}; Swift data from \citet{Hinkle2021b} and optical data from \citet{Hung2020,Holoien2020,Holoien2019,Holoien2016} for AT~2018hyz, AT~2018dyb, AT~2018zr and ASASSN-14li respectively. The listed epoch is with respect to peak, except for ASASSN-14li, where it is with respect to the first detection.
      }
\label{fig:Jiang_comparisons}
\end{figure}

%\tr{Should say more here about viewing angles and how many TDEs should look like this?}
There is a small sample of TDEs in the literature that have been observed close to the peak in the IR. All of these that we know of come from the NEOWISE survey, with no peak NIR observations previously reported. In the upper panel of Fig. \ref{fig:Jiang_comparisons} we compare the W1 photometry from AT~2019azh with the sample of NEOWISE observed TDEs presented in \citet{Jiang2021b}. The peak luminosity of AT~2019azh is quite typical for this sample of objects and the light curve evolution strongly resembles the TDE AT~2018dyb \citep[also known as ASASSN-18pg;][]{Holoien2020,Leloudas2019}, as well as being quite similar to ASASSN-14li and AT~2018hyz, with slight offsets in luminosity. There are only four TDEs in this sample that have MIR observations within 30~d of their peak; however, three of these exhibit a very similar MIR photometric evolution as AT~2019azh, with a sharp decline between the first observation and the later observations. It follows that, similarly to AT~2019azh, the peak MIR emission could be arising from direct TDE emission rather than emission from dust in these TDEs.% despite the fact that these IR detections have always previously been interpreted as IR echoes. 

To further explore this, we show the SEDs of these TDEs at the epoch of the first MIR observation in the lower panel Fig. \ref{fig:Jiang_comparisons}. The figure shows that the SED of AT~2018dyb is remarkably similar to that of AT~2019azh at an almost identical epoch, in both the UVO and MIR regions. In particular, the W1-W2 colour is similar to that of AT~2019azh. This indicates that a reprocessing scenario could potentially explain the IR excess seen in this TDE. %AT~2018dyb was not detected in X-rays in 54 stacked Swift observations covering the initial $\sim 300$d of the TDE evolution \cite{Holoien2020}, which could indicate a greater degree of obscuration for this TDE. In the viewing angle dependent model, this requires a very steep viewing angle in order to absorb the X-rays at late times. %\tr{Need more specific upper limits to compare. Stacking all Swift data gives a tentative 2.7 sigma detection.}.
The other TDEs show variety, with both AT~2018hyz and AT~2018zr being less luminous in the UV, indicating a cooler and more thermal spectrum, or potentially host dust extinction, while ASASSN-14li appears both significantly less luminous and displays a U-shaped SED that is indicative of hot dust emission, as claimed in \citet{Jiang2016}. Furthermore, the $W1-W2$ colours of AT~2018zr and AT~2018hyz are also redder than those of the other TDEs. These TDEs are additionally all observed later than 20~d, in contrast with AT~2019azh and AT~2018dyb which are observed at $\sim$7~d, preventing a direct comparison. As we lack the NIR data to constrain the wavelength region between 1-2 $\mu$m, it is difficult to conclusively determine the origin of the IR emission in these cases. %There are additional TDE candidates, such as extreme coronal line emitters , ambiguous nuclear transients and bowen fluorescence flares which, due to their slower evolution, have MIR data available around peak. However, these objects show much more luminous and long lived IR echoes

\section{Conclusions}
\label{sec:conclusions}
In this work, we have presented IR observations of the TDE AT~2019azh and examined whether they are consistent with emission from dust heated by the TDE, or direct emission from the TDE in a reprocessing scenario. We summarise our findings as follows:

\begin{enumerate}

\item AT~2019azh exhibits a significant IR excess compared to a blackbody fit to the UVO data. The excess in the NIR alone before 40~d can be fit by a blackbody corresponding to very hot (>2000~K) graphite dust, but the NIR and MIR data together can not be consistently fit.

\item The IR excess exhibits a power law shape, as predicted for non-thermal emission arising from free-free opacity in the IR in TDEs. This requires the presence of a dense reprocessing envelope.

\item The early time UVO+IR SED can be well-reproduced by spectral models produced from simulations of emission from a super-Eddington accretion flow. This model is viewing-angle dependent, and by comparing the SED to these models, we find that a viewing angle of $\sim$60\textdegree~is necessary to reproduce the data. The implied mass accretion rate indicates that the disrupted star had a mass $M_{\star}>2$M$_{\odot}$.

\item The X-ray brightening of AT~2019azh supports a reprocessing scenario, with either the viewing-angle dependent model or the CIO model qualitatively reproducing the observations.

\item Late time IR data show the presence of an IR echo from relatively cold (<1000~K) dust. We model the IR echo and show that the dust must be distant from the TDE (>0.65~pc) to have these low temperatures, and must have a relatively small geometrical covering factor in order to consistently produce an IR echo with a short timescale. We find evidence for a clumpy distribution of circumnuclear clouds, which is consistent with evidence for a similar distribution observed in the radio, although probing different distances from the SMBH. There is no evidence for high temperature dust associated with dust sublimation and therefore the dust distribution is independent of the TDE flare.

\item Considering the small sample of previous TDEs observed close to the peak in the IR, we find that AT~2018dyb potentially also exhibited an IR excess due to non-thermal TDE emission.

\end{enumerate}

Further observations of TDEs in the IR will be able to measure the relative rates of TDEs that show such a non-thermal IR excess, and determine whether these are consistent with the expected distribution of viewing angles for TDEs. Furthermore, multi-wavelength observations of TDEs that include the IR will be able to produce observables that can determine which reprocessing model is responsible for the TDEs we observe, and we invite further modelling efforts to capitalise on this. As in the case of AT~2019azh, observing both in the NIR \textit{and} MIR will be crucial for distinguishing between non-thermal emission, and dust emission. Given that the NEOWISE survey has ended, future MIR facilities such as the NEO Surveyor \citep{Mainzer2023} and SPHEREx \citep{sphereX} will be essential.

\begin{acknowledgements}

T.M.R. thanks Thomas Wevers, Yanan Wang, Iair Arcavi and Peter Clark for helpful comments during preperation of the manuscript. T.M.R is part of the Cosmic Dawn Center (DAWN), which is funded by the Danish National Research Foundation under grant DNRF140. T.M.R and S. Mattila acknowledge support from the Research Council of Finland project 350458. S. Moran is funded by Leverhulme Trust grant RPG-2023-240. LT and LD acknowledge support from the National Natural Science Foundation of China and the Hong Kong Research Grants Council (N\_HKU782/23, HKU17305124). CPG acknowledges financial support from the Secretary of Universities and Research (Government of Catalonia) and by the Horizon 2020 Research and Innovation Programme of the European Union under the Marie Sk\l{}odowska-Curie and the Beatriu de Pin\'os 2021 BP 00168 programme, from the Spanish Ministerio de Ciencia e Innovaci\'on (MCIN) and the Agencia Estatal de Investigaci\'on (AEI) 10.13039/501100011033 under the PID2023-151307NB-I00 SNNEXT project, from Centro Superior de Investigaciones Cient\'ificas  (CSIC) under the PIE project 20215AT016 and the program Unidad de Excelencia Mar\'ia  de Maeztu CEX2020-001058-M, and from the Departament de Recerca i Universitats de la Generalitat de Catalunya through the 2021-SGR-01270 grant.FEB acknowledges support from ANID-Chile BASAL CATA FB210003, FONDECYT Regular 1241005, and Millennium Science Initiative, AIM23-0001.

%Data
% ESO
This work is partially based on observations collected at the European Southern Observatory under ESO programme 199.D-0143(Q).
% NOT
This research was partly based on observations made with the Nordic Optical Telescope (program IDs: P59-506 \& P68-505) owned in collaboration by the University of Turku and Aarhus University, and operated jointly by Aarhus University, the University of Turku and the University of Oslo, representing Denmark, Finland and Norway, the University of Iceland and Stockholm University at the Observatorio del Roque de los Muchachos, La Palma, Spain, of the Instituto de Astrofisica de Canarias.
% 2MASS
This publication makes use of data products from the Two Micron All Sky Survey, which is a joint project of the University of Massachusetts and the Infrared Processing and Analysis Center/California Institute of Technology, funded by the National Aeronautics and Space Administration and the National Science Foundation.
% NEOWISE
This publication makes use of data products from the Near-Earth Object Wide-field Infrared Survey Explorer (NEOWISE), which is a joint project of the Jet Propulsion Laboratory/California Institute of Technology and the University of California, Los Angeles. NEOWISE is funded by the National Aeronautics and Space Administration.
%NED, python packages.
% Astropy
This work made use of Astropy: a community-developed core Python package and an ecosystem of tools and resources for astronomy \citep[Astropy Collaboration][]{astropy:2013,astropy:2018,astropy:2022}.

\end{acknowledgements}

% WARNING
%-------------------------------------------------------------------
% Please note that we have included the references to the file aa.dem in
% order to compile it, but we ask you to:
%
% - use BibTeX with the regular commands:
%   \bibliographystyle{aa} % style aa.bst
%   \bibliography{Yourfile} % your references Yourfile.bib
%
% - join the .bib files when you upload your source files
%-------------------------------------------------------------------

\bibliographystyle{aa} % style aa.bst
\bibliography{bibliography} % your references Yourfile.bib

\begin{thebibliography}{85}
\expandafter\ifx\csname natexlab\endcsname\relax\def\natexlab#1{#1}\fi

\bibitem[{{Alard} \& {Lupton}(1998)}]{Alard1998}
{Alard}, C. \& {Lupton}, R.~H. 1998, \apj, 503, 325

\bibitem[{Alibay {et~al.}(2023)Alibay, Sindiy, Jansma, Reynerson, Rice, Rocca,
  Susca, Unwin, Akeson, Mihaly, \& Werner}]{sphereX}
Alibay, F., Sindiy, O.~V., Jansma, P. A.~T., {et~al.} 2023, in 2023 IEEE
  Aerospace Conference, 1--18

\bibitem[{{Almeida} {et~al.}(2023){Almeida}, {Anderson},
  {Argudo-Fern{\'a}ndez}, {Badenes}, {Barger}, {Barrera-Ballesteros}, {Bender},
  {Benitez}, {Besser}, {Bird}, {Bizyaev}, {Blanton}, {Bochanski}, {Bovy},
  {Brandt}, {Brownstein}, {Buchner}, {Bulbul}, {Burchett}, {Cano D{\'\i}az},
  {Carlberg}, {Casey}, {Chandra}, {Cherinka}, {Chiappini}, {Coker}, {Comparat},
  {Conroy}, {Contardo}, {Cortes}, {Covey}, {Crane}, {Cunha}, {Dabbieri},
  {Davidson}, {Davis}, {de Andrade Queiroz}, {De Lee}, {M{\'e}ndez Delgado},
  {Demasi}, {Di Mille}, {Donor}, {Dow}, {Dwelly}, {Eracleous}, {Eriksen},
  {Fan}, {Farr}, {Frederick}, {Fries}, {Frinchaboy}, {G{\"a}nsicke}, {Ge},
  {Gonz{\'a}lez {\'A}vila}, {Grabowski}, {Grier}, {Guiglion}, {Gupta}, {Hall},
  {Hawkins}, {Hayes}, {Hermes}, {Hern{\'a}ndez-Garc{\'\i}a}, {Hogg},
  {Holtzman}, {Ibarra-Medel}, {Ji}, {Jofre}, {Johnson}, {Jones}, {Kinemuchi},
  {Kluge}, {Koekemoer}, {Kollmeier}, {Kounkel}, {Krishnarao}, {Krumpe},
  {Lacerna}, {Lago}, {Laporte}, {Liu}, {Liu}, {Liu}, {Lopes}, {Macktoobian},
  {Majewski}, {Malanushenko}, {Maoz}, {Masseron}, {Masters}, {Matijevic},
  {McBride}, {Medan}, {Merloni}, {Morrison}, {Myers}, {M{\'e}sz{\'a}ros},
  {Negrete}, {Nidever}, {Nitschelm}, {Oravetz}, {Oravetz}, {Pan}, {Peng},
  {Pinsonneault}, {Pogge}, {Qiu}, {Ramirez}, {Rix}, {Fern{\'a}ndez Rosso},
  {Runnoe}, {Salvato}, {Sanchez}, {Santana}, {Saydjari}, {Sayres},
  {Schlaufman}, {Schneider}, {Schwope}, {Serna}, {Shen}, {Sobeck}, {Song},
  {Souto}, {Spoo}, {Stassun}, {Steinmetz}, {Straumit}, {Stringfellow},
  {S{\'a}nchez-Gallego}, {Taghizadeh-Popp}, {Tayar}, {Thakar}, {Tissera},
  {Tkachenko}, {Hernandez Toledo}, {Trakhtenbrot}, {Fern{\'a}ndez-Trincado},
  {Troup}, {Trump}, {Tuttle}, {Ulloa}, {Vazquez-Mata}, {Vera Alfaro},
  {Villanova}, {Wachter}, {Weijmans}, {Wheeler}, {Wilson}, {Wojno}, {Wolf},
  {Xue}, {Ybarra}, {Zari}, \& {Zasowski}}]{Almeida2023}
{Almeida}, A., {Anderson}, S.~F., {Argudo-Fern{\'a}ndez}, M., {et~al.} 2023,
  \apjs, 267, 44

\bibitem[{{Arcavi} {et~al.}(2014){Arcavi}, {Gal-Yam}, {Sullivan}, {Pan},
  {Cenko}, {Horesh}, {Ofek}, {De Cia}, {Yan}, {Yang}, {Howell}, {Tal},
  {Kulkarni}, {Tendulkar}, {Tang}, {Xu}, {Sternberg}, {Cohen}, {Bloom},
  {Nugent}, {Kasliwal}, {Perley}, {Quimby}, {Miller}, {Theissen}, \&
  {Laher}}]{Arcavi2014}
{Arcavi}, I., {Gal-Yam}, A., {Sullivan}, M., {et~al.} 2014, \apj, 793, 38

\bibitem[{{Astropy Collaboration} {et~al.}(2022){Astropy Collaboration},
  {Price-Whelan}, {Lim}, {Earl}, {Starkman}, {Bradley}, {Shupe}, {Patil},
  {Corrales}, {Brasseur}, {N{\"o}the}, {Donath}, {Tollerud}, {Morris},
  {Ginsburg}, {Vaher}, {Weaver}, {Tocknell}, {Jamieson}, {van Kerkwijk},
  {Robitaille}, {Merry}, {Bachetti}, {G{\"u}nther}, {Aldcroft},
  {Alvarado-Montes}, {Archibald}, {B{\'o}di}, {Bapat}, {Barentsen},
  {Baz{\'a}n}, {Biswas}, {Boquien}, {Burke}, {Cara}, {Cara}, {Conroy},
  {Conseil}, {Craig}, {Cross}, {Cruz}, {D'Eugenio}, {Dencheva}, {Devillepoix},
  {Dietrich}, {Eigenbrot}, {Erben}, {Ferreira}, {Foreman-Mackey}, {Fox},
  {Freij}, {Garg}, {Geda}, {Glattly}, {Gondhalekar}, {Gordon}, {Grant},
  {Greenfield}, {Groener}, {Guest}, {Gurovich}, {Handberg}, {Hart},
  {Hatfield-Dodds}, {Homeier}, {Hosseinzadeh}, {Jenness}, {Jones}, {Joseph},
  {Kalmbach}, {Karamehmetoglu}, {Ka{\l}uszy{\'n}ski}, {Kelley}, {Kern},
  {Kerzendorf}, {Koch}, {Kulumani}, {Lee}, {Ly}, {Ma}, {MacBride}, {Maljaars},
  {Muna}, {Murphy}, {Norman}, {O'Steen}, {Oman}, {Pacifici}, {Pascual},
  {Pascual-Granado}, {Patil}, {Perren}, {Pickering}, {Rastogi}, {Roulston},
  {Ryan}, {Rykoff}, {Sabater}, {Sakurikar}, {Salgado}, {Sanghi}, {Saunders},
  {Savchenko}, {Schwardt}, {Seifert-Eckert}, {Shih}, {Jain}, {Shukla}, {Sick},
  {Simpson}, {Singanamalla}, {Singer}, {Singhal}, {Sinha}, {Sip{\H{o}}cz},
  {Spitler}, {Stansby}, {Streicher}, {{\v{S}}umak}, {Swinbank}, {Taranu},
  {Tewary}, {Tremblay}, {Val-Borro}, {Van Kooten}, {Vasovi{\'c}}, {Verma}, {de
  Miranda Cardoso}, {Williams}, {Wilson}, {Winkel}, {Wood-Vasey}, {Xue},
  {Yoachim}, {Zhang}, {Zonca}, \& {Astropy Project
  Contributors}}]{astropy:2022}
{Astropy Collaboration}, {Price-Whelan}, A.~M., {Lim}, P.~L., {et~al.} 2022,
  \apj, 935, 167

\bibitem[{{Astropy Collaboration} {et~al.}(2018){Astropy Collaboration},
  {Price-Whelan}, {Sip{\H o}cz}, {G{\"u}nther}, {Lim}, {Crawford}, {Conseil},
  {Shupe}, {Craig}, {Dencheva}, {Ginsburg}, {VanderPlas}, {Bradley},
  {P{\'e}rez-Su{\'a}rez}, {de Val-Borro}, {Paper Contributors}, {Aldcroft},
  {Cruz}, {Robitaille}, {Tollerud}, {Coordination Committee}, {Ardelean},
  {Babej}, {Bach}, {Bachetti}, {Bakanov}, {Bamford}, {Barentsen}, {Barmby},
  {Baumbach}, {Berry}, {Biscani}, {Boquien}, {Bostroem}, {Bouma}, {Brammer},
  {Bray}, {Breytenbach}, {Buddelmeijer}, {Burke}, {Calderone}, {Cano
  Rodr{\'{\i}}guez}, {Cara}, {Cardoso}, {Cheedella}, {Copin}, {Corrales},
  {Crichton}, {D{\'A}vella}, {Deil}, {Depagne}, {Dietrich}, {Donath},
  {Droettboom}, {Earl}, {Erben}, {Fabbro}, {Ferreira}, {Finethy}, {Fox},
  {Garrison}, {Gibbons}, {Goldstein}, {Gommers}, {Greco}, {Greenfield},
  {Groener}, {Grollier}, {Hagen}, {Hirst}, {Homeier}, {Horton}, {Hosseinzadeh},
  {Hu}, {Hunkeler}, {Ivezi{\'c}}, {Jain}, {Jenness}, {Kanarek}, {Kendrew},
  {Kern}, {Kerzendorf}, {Khvalko}, {King}, {Kirkby}, {Kulkarni}, {Kumar},
  {Lee}, {Lenz}, {Littlefair}, {Ma}, {Macleod}, {Mastropietro}, {McCully},
  {Montagnac}, {Morris}, {Mueller}, {Mumford}, {Muna}, {Murphy}, {Nelson},
  {Nguyen}, {Ninan}, {N{\"o}the}, {Ogaz}, {Oh}, {Parejko}, {Parley}, {Pascual},
  {Patil}, {Patil}, {Plunkett}, {Prochaska}, {Rastogi}, {Reddy Janga},
  {Sabater}, {Sakurikar}, {Seifert}, {Sherbert}, {Sherwood-Taylor}, {Shih},
  {Sick}, {Silbiger}, {Singanamalla}, {Singer}, {Sladen}, {Sooley},
  {Sornarajah}, {Streicher}, {Teuben}, {Thomas}, {Tremblay}, {Turner},
  {Terr{\'o}n}, {van Kerkwijk}, {de la Vega}, {Watkins}, {Weaver}, {Whitmore},
  {Woillez}, {Zabalza}, \& {Contributors}}]{astropy:2018}
{Astropy Collaboration}, {Price-Whelan}, A.~M., {Sip{\H o}cz}, B.~M., {et~al.}
  2018, \aj, 156, 123

\bibitem[{{Astropy Collaboration} {et~al.}(2013){Astropy Collaboration},
  {Robitaille}, {Tollerud}, {Greenfield}, {Droettboom}, {Bray}, {Aldcroft},
  {Davis}, {Ginsburg}, {Price-Whelan}, {Kerzendorf}, {Conley}, {Crighton},
  {Barbary}, {Muna}, {Ferguson}, {Grollier}, {Parikh}, {Nair}, {Unther},
  {Deil}, {Woillez}, {Conseil}, {Kramer}, {Turner}, {Singer}, {Fox}, {Weaver},
  {Zabalza}, {Edwards}, {Azalee Bostroem}, {Burke}, {Casey}, {Crawford},
  {Dencheva}, {Ely}, {Jenness}, {Labrie}, {Lian Lim}, {Pierfederici},
  {Pontzen}, {Ptak}, {Refsdal}, {Servillat}, \& {Streicher}}]{astropy:2013}
{Astropy Collaboration}, {Robitaille}, T.~P., {Tollerud}, E.~J., {et~al.} 2013,
  \aap, 558, A33

\bibitem[{{Baskin} \& {Laor}(2018)}]{Baskin2018}
{Baskin}, A. \& {Laor}, A. 2018, \mnras, 474, 1970

\bibitem[{{Bennett} {et~al.}(2014){Bennett}, {Larson}, {Weiland}, \&
  {Hinshaw}}]{Bennett2014}
{Bennett}, C.~L., {Larson}, D., {Weiland}, J.~L., \& {Hinshaw}, G. 2014, \apj,
  794, 135

\bibitem[{{Bonnerot} {et~al.}(2021){Bonnerot}, {Lu}, \&
  {Hopkins}}]{Bonnerot2021}
{Bonnerot}, C., {Lu}, W., \& {Hopkins}, P.~F. 2021, \mnras, 504, 4885

\bibitem[{{Brennan} \& {Fraser}(2022)}]{Brennan2022}
{Brennan}, S.~J. \& {Fraser}, M. 2022, \aap, 667, A62

\bibitem[{{Christopher} {et~al.}(2005){Christopher}, {Scoville}, {Stolovy}, \&
  {Yun}}]{Christopher2005}
{Christopher}, M.~H., {Scoville}, N.~Z., {Stolovy}, S.~R., \& {Yun}, M.~S.
  2005, \apj, 622, 346

\bibitem[{{Dai} {et~al.}(2018){Dai}, {McKinney}, {Roth}, {Ramirez-Ruiz}, \&
  {Miller}}]{Dai2018}
{Dai}, L., {McKinney}, J.~C., {Roth}, N., {Ramirez-Ruiz}, E., \& {Miller},
  M.~C. 2018, \apjl, 859, L20

\bibitem[{{Donley} {et~al.}(2002){Donley}, {Brandt}, {Eracleous}, \&
  {Boller}}]{Donley2002}
{Donley}, J.~L., {Brandt}, W.~N., {Eracleous}, M., \& {Boller}, T. 2002, \aj,
  124, 1308

\bibitem[{{Dou} {et~al.}(2016){Dou}, {Wang}, {Jiang}, {Yang}, {Lyu}, \&
  {Zhou}}]{Dou2016}
{Dou}, L., {Wang}, T.-g., {Jiang}, N., {et~al.} 2016, \apj, 832, 188

\bibitem[{{Dwek}(1983)}]{dwek1983}
{Dwek}, E. 1983, \apj, 274, 175

\bibitem[{{Evans} \& {Kochanek}(1989)}]{Evans1989}
{Evans}, C.~R. \& {Kochanek}, C.~S. 1989, \apjl, 346, L13

\bibitem[{{Faris} {et~al.}(2024){Faris}, {Arcavi}, {Makrygianni}, {Hiramatsu},
  {Terreran}, {Farah}, {Howell}, {McCully}, {Newsome}, {Padilla Gonzalez},
  {Pellegrino}, {Bostroem}, {Abojanb}, {Lam}, {Tomasella}, {Brink},
  {Filippenko}, {French}, {Clark}, {Graur}, {Leloudas}, {Gromadzki},
  {Anderson}, {Nicholl}, {Guti{\'e}rrez}, {Kankare}, {Inserra}, {Galbany},
  {Reynolds}, {Mattila}, {Heikkil{\"a}}, {Wang}, {Onori}, {Wevers}, {Coughlin},
  {Charalampopoulos}, \& {Johansson}}]{Faris2023}
{Faris}, S., {Arcavi}, I., {Makrygianni}, L., {et~al.} 2024, \apj, 969, 104

\bibitem[{{Fitzpatrick}(1999)}]{Fitzpatrick1999}
{Fitzpatrick}, E.~L. 1999, \pasp, 111, 63

\bibitem[{{Foreman-Mackey} {et~al.}(2013){Foreman-Mackey}, {Hogg}, {Lang}, \&
  {Goodman}}]{ForemanMackey2013}
{Foreman-Mackey}, D., {Hogg}, D.~W., {Lang}, D., \& {Goodman}, J. 2013, \pasp,
  125, 306

\bibitem[{{French} {et~al.}(2023){French}, {Earl}, {Novack}, {Pardasani},
  {Pillai}, {Tripathi}, \& {Verrico}}]{French2023}
{French}, K.~D., {Earl}, N., {Novack}, A.~B., {et~al.} 2023, \apj, 950, 153

\bibitem[{{French} {et~al.}(2020){French}, {Wevers}, {Law-Smith}, {Graur}, \&
  {Zabludoff}}]{French2020}
{French}, K.~D., {Wevers}, T., {Law-Smith}, J., {Graur}, O., \& {Zabludoff},
  A.~I. 2020, \ssr, 216, 32

\bibitem[{{Gezari}(2021)}]{Gezari2021}
{Gezari}, S. 2021, \araa, 59, 21

\bibitem[{{Goodwin} {et~al.}(2022){Goodwin}, {van Velzen}, {Miller-Jones},
  {Mummery}, {Bietenholz}, {Wederfoort}, {Hammerstein}, {Bonnerot}, {Hoffmann},
  \& {Yan}}]{Goodwin2022}
{Goodwin}, A.~J., {van Velzen}, S., {Miller-Jones}, J.~C.~A., {et~al.} 2022,
  \mnras, 511, 5328

\bibitem[{{Graham} {et~al.}(1983){Graham}, {Meikle}, {Selby}, {Allen}, {Evans},
  {Pearce}, {Bode}, {Longmore}, \& {Williams}}]{graham1983}
{Graham}, J.~R., {Meikle}, W.~P.~S., {Selby}, M.~J., {et~al.} 1983, \nat, 304,
  709

\bibitem[{{Guhathakurta} \& {Draine}(1989)}]{Guhathakurta1989}
{Guhathakurta}, P. \& {Draine}, B.~T. 1989, \apj, 345, 230

\bibitem[{{Guillochon} {et~al.}(2014){Guillochon}, {Manukian}, \&
  {Ramirez-Ruiz}}]{Guillochon2014}
{Guillochon}, J., {Manukian}, H., \& {Ramirez-Ruiz}, E. 2014, \apj, 783, 23

\bibitem[{{Guillochon} {et~al.}(2018){Guillochon}, {Nicholl}, {Villar},
  {Mockler}, {Narayan}, {Mandel}, {Berger}, \& {Williams}}]{Guillochon2018}
{Guillochon}, J., {Nicholl}, M., {Villar}, V.~A., {et~al.} 2018, \apjs, 236, 6

\bibitem[{{Guillochon} \& {Ramirez-Ruiz}(2013)}]{Guillochon2013}
{Guillochon}, J. \& {Ramirez-Ruiz}, E. 2013, \apj, 767, 25

\bibitem[{{Guolo} {et~al.}(2024){Guolo}, {Gezari}, {Yao}, {van Velzen},
  {Hammerstein}, {Cenko}, \& {Tokayer}}]{Guolo2024}
{Guolo}, M., {Gezari}, S., {Yao}, Y., {et~al.} 2024, \apj, 966, 160

\bibitem[{{Hammerstein} {et~al.}(2023){Hammerstein}, {van Velzen}, {Gezari},
  {Cenko}, {Yao}, {Ward}, {Frederick}, {Villanueva}, {Somalwar}, {Graham},
  {Kulkarni}, {Stern}, {Andreoni}, {Bellm}, {Dekany}, {Dhawan}, {Drake},
  {Fremling}, {Gatkine}, {Groom}, {Ho}, {Kasliwal}, {Karambelkar}, {Kool},
  {Masci}, {Medford}, {Perley}, {Purdum}, {van Roestel}, {Sharma}, {Sollerman},
  {Taggart}, \& {Yan}}]{Hammerstein2023}
{Hammerstein}, E., {van Velzen}, S., {Gezari}, S., {et~al.} 2023, \apj, 942, 9

\bibitem[{{Hinkle} {et~al.}(2021{\natexlab{a}}){Hinkle}, {Holoien}, {Auchettl},
  {Shappee}, {Neustadt}, {Payne}, {Brown}, {Kochanek}, {Stanek}, {Graham},
  {Tucker}, {Do}, {Anderson}, {Bose}, {Chen}, {Coulter}, {Dimitriadis}, {Dong},
  {Foley}, {Huber}, {Hung}, {Kilpatrick}, {Pignata}, {Piro}, {Rojas-Bravo},
  {Siebert}, {Stalder}, {Thompson}, {Tonry}, {Vallely}, \&
  {Wisniewski}}]{Hinkle2021a}
{Hinkle}, J.~T., {Holoien}, T.~W.~S., {Auchettl}, K., {et~al.}
  2021{\natexlab{a}}, \mnras, 500, 1673

\bibitem[{{Hinkle} {et~al.}(2021{\natexlab{b}}){Hinkle}, {Holoien}, {Shappee},
  \& {Auchettl}}]{Hinkle2021b}
{Hinkle}, J.~T., {Holoien}, T. W.~S., {Shappee}, B.~J., \& {Auchettl}, K.
  2021{\natexlab{b}}, \apj, 910, 83

\bibitem[{{Holoien} {et~al.}(2020){Holoien}, {Auchettl}, {Tucker}, {Shappee},
  {Patel}, {Miller-Jones}, {Mockler}, {Groenewald}, {Hinkle}, {Brown},
  {Kochanek}, {Stanek}, {Chen}, {Dong}, {Prieto}, {Thompson}, {Beaton},
  {Connor}, {Cowperthwaite}, {Dahmen}, {French}, {Morrell}, {Buckley},
  {Gromadzki}, {Roy}, {Coulter}, {Dimitriadis}, {Foley}, {Kilpatrick}, {Piro},
  {Rojas-Bravo}, {Siebert}, \& {van Velzen}}]{Holoien2020}
{Holoien}, T. W.~S., {Auchettl}, K., {Tucker}, M.~A., {et~al.} 2020, \apj, 898,
  161

\bibitem[{{Holoien} {et~al.}(2019){Holoien}, {Huber}, {Shappee}, {Eracleous},
  {Auchettl}, {Brown}, {Tucker}, {Chambers}, {Kochanek}, {Stanek}, {Rest},
  {Bersier}, {Post}, {Aldering}, {Ponder}, {Simon}, {Kankare}, {Dong},
  {Hallinan}, {Reddy}, {Sanders}, {Topping}, {Pan-STARRS}, {Bulger}, {Lowe},
  {Magnier}, {Schultz}, {Waters}, {Willman}, {Wright}, {Young}, {ASAS-SN},
  {Dong}, {Prieto}, {Thompson}, {ATLAS}, {Denneau}, {Flewelling}, {Heinze},
  {Smartt}, {Smith}, {Stalder}, {Tonry}, \& {Weiland}}]{Holoien2019}
{Holoien}, T.~W.~S., {Huber}, M.~E., {Shappee}, B.~J., {et~al.} 2019, \apj,
  880, 120

\bibitem[{{Holoien} {et~al.}(2016){Holoien}, {Kochanek}, {Prieto}, {Stanek},
  {Dong}, {Shappee}, {Grupe}, {Brown}, {Basu}, {Beacom}, {Bersier},
  {Brimacombe}, {Danilet}, {Falco}, {Guo}, {Jose}, {Herczeg}, {Long},
  {Pojmanski}, {Simonian}, {Szczygie{\l}}, {Thompson}, {Thorstensen}, {Wagner},
  \& {Wo{\'z}niak}}]{Holoien2016}
{Holoien}, T.~W.~S., {Kochanek}, C.~S., {Prieto}, J.~L., {et~al.} 2016, \mnras,
  455, 2918

\bibitem[{{Hung} {et~al.}(2020){Hung}, {Foley}, {Ramirez-Ruiz}, {Dai},
  {Auchettl}, {Kilpatrick}, {Mockler}, {Brown}, {Coulter}, {Dimitriadis},
  {Holoien}, {Law-Smith}, {Piro}, {Rest}, {Rojas-Bravo}, \&
  {Siebert}}]{Hung2020}
{Hung}, T., {Foley}, R.~J., {Ramirez-Ruiz}, E., {et~al.} 2020, \apj, 903, 31

\bibitem[{{Jiang} {et~al.}(2016){Jiang}, {Dou}, {Wang}, {Yang}, {Lyu}, \&
  {Zhou}}]{Jiang2016}
{Jiang}, N., {Dou}, L., {Wang}, T., {et~al.} 2016, \apjl, 828, L14

\bibitem[{{Jiang} {et~al.}(2021{\natexlab{a}}){Jiang}, {Wang}, {Dou}, {Shu},
  {Hu}, {Liu}, {Wang}, {Yan}, {Sheng}, {Yang}, {Sun}, \& {Zhou}}]{Jiang2021a}
{Jiang}, N., {Wang}, T., {Dou}, L., {et~al.} 2021{\natexlab{a}}, \apjs, 252, 32

\bibitem[{{Jiang} {et~al.}(2021{\natexlab{b}}){Jiang}, {Wang}, {Hu}, {Sun},
  {Dou}, \& {Xiao}}]{Jiang2021b}
{Jiang}, N., {Wang}, T., {Hu}, X., {et~al.} 2021{\natexlab{b}}, \apj, 911, 31

\bibitem[{{Kasen} {et~al.}(2006){Kasen}, {Thomas}, \& {Nugent}}]{Kasen2006}
{Kasen}, D., {Thomas}, R.~C., \& {Nugent}, P. 2006, \apj, 651, 366

\bibitem[{{Kippenhahn} {et~al.}(2013){Kippenhahn}, {Weigert}, \&
  {Weiss}}]{Kippenhahn2013}
{Kippenhahn}, R., {Weigert}, A., \& {Weiss}, A. 2013, {Stellar Structure and
  Evolution} (Springer Berlin, Heidelberg)

\bibitem[{{Kool} {et~al.}(2020){Kool}, {Reynolds}, {Mattila}, {Kankare},
  {P{\'e}rez-Torres}, {Efstathiou}, {Ryder}, {Romero-Ca{\~n}izales}, {Lu},
  {Heikkil{\"a}}, {Anderson}, {Berton}, {Bright}, {Cannizzaro}, {Eappachen},
  {Fraser}, {Gromadzki}, {Jonker}, {Kuncarayakti}, {Lundqvist}, {Maeda},
  {McDermid}, {Medling}, {Moran}, {Reguitti}, {Shahbandeh}, {Tsygankov}, {U},
  \& {Wevers}}]{kool2020}
{Kool}, E.~C., {Reynolds}, T.~M., {Mattila}, S., {et~al.} 2020, \mnras, 498,
  2167

\bibitem[{{Lang}(2014)}]{Lang2014}
{Lang}, D. 2014, \aj, 147, 108

\bibitem[{{Leloudas} {et~al.}(2019){Leloudas}, {Dai}, {Arcavi}, {Vreeswijk},
  {Mockler}, {Roy}, {Malesani}, {Schulze}, {Wevers}, {Fraser}, {Ramirez-Ruiz},
  {Auchettl}, {Burke}, {Cannizzaro}, {Charalampopoulos}, {Chen}, {Cikota},
  {Della Valle}, {Galbany}, {Gromadzki}, {Heintz}, {Hiramatsu}, {Jonker},
  {Kostrzewa-Rutkowska}, {Maguire}, {Mandel}, {Nicholl}, {Onori}, {Roth},
  {Smartt}, {Wyrzykowski}, \& {Young}}]{Leloudas2019}
{Leloudas}, G., {Dai}, L., {Arcavi}, I., {et~al.} 2019, \apj, 887, 218

\bibitem[{{Liu} {et~al.}(2022){Liu}, {Dou}, {Chen}, \& {Shen}}]{Liu2022}
{Liu}, X.-L., {Dou}, L.-M., {Chen}, J.-H., \& {Shen}, R.-F. 2022, \apj, 925, 67

\bibitem[{{Loeb} \& {Ulmer}(1997)}]{Loeb1997}
{Loeb}, A. \& {Ulmer}, A. 1997, \apj, 489, 573

\bibitem[{{Lu} \& {Bonnerot}(2020)}]{Lu2020}
{Lu}, W. \& {Bonnerot}, C. 2020, \mnras, 492, 686

\bibitem[{{Lu} {et~al.}(2016){Lu}, {Kumar}, \& {Evans}}]{Lu2016}
{Lu}, W., {Kumar}, P., \& {Evans}, N.~J. 2016, \mnras, 458, 575

\bibitem[{{Maeda} {et~al.}(2015){Maeda}, {Nozawa}, {Nagao}, \&
  {Motohara}}]{Maeda2015}
{Maeda}, K., {Nozawa}, T., {Nagao}, T., \& {Motohara}, K. 2015, \mnras, 452,
  3281

\bibitem[{{Mainzer} {et~al.}(2014){Mainzer}, {Bauer}, {Cutri}, {Grav},
  {Masiero}, {Beck}, {Clarkson}, {Conrow}, {Dailey}, {Eisenhardt}, {Fabinsky},
  {Fajardo-Acosta}, {Fowler}, {Gelino}, {Grillmair}, {Heinrichsen}, {Kendall},
  {Kirkpatrick}, {Liu}, {Masci}, {McCallon}, {Nugent}, {Papin}, {Rice},
  {Royer}, {Ryan}, {Sevilla}, {Sonnett}, {Stevenson}, {Thompson}, {Wheelock},
  {Wiemer}, {Wittman}, {Wright}, \& {Yan}}]{Mainzer2014}
{Mainzer}, A., {Bauer}, J., {Cutri}, R.~M., {et~al.} 2014, \apj, 792, 30

\bibitem[{{Mainzer} {et~al.}(2023){Mainzer}, {Masiero}, {Abell}, {Bauer},
  {Bottke}, {Buratti}, {Carey}, {Cotto-Figueroa}, {Cutri}, {Dahlen},
  {Eisenhardt}, {Fernandez}, {Furfaro}, {Grav}, {Hoffman}, {Kelley}, {Kim},
  {Kirkpatrick}, {Lawler}, {Lilly}, {Liu}, {Marocco}, {Marsh}, {Masci},
  {McMurtry}, {Pourrahmani}, {Reinhart}, {Ressler}, {Satpathy}, {Schambeau},
  {Sonnett}, {Spahr}, {Surace}, {Vaquero}, {Wright}, {Zengilowski}, \& {NEO
  Surveyor Mission Team}}]{Mainzer2023}
{Mainzer}, A.~K., {Masiero}, J.~R., {Abell}, P.~A., {et~al.} 2023, Planet. Sci.
  J., 4, 224

\bibitem[{{Masterson} {et~al.}(2024){Masterson}, {De}, {Panagiotou}, {Kara},
  {Arcavi}, {Eilers}, {Frostig}, {Gezari}, {Grotova}, {Liu}, {Malyali},
  {Meisner}, {Merloni}, {Newsome}, {Rau}, {Simcoe}, \& {van
  Velzen}}]{masterson2024}
{Masterson}, M., {De}, K., {Panagiotou}, C., {et~al.} 2024, \apj, 961, 211

\bibitem[{{Masterson} {et~al.}(2025){Masterson}, {De}, {Panagiotou}, {Kara},
  {Eilers}, {Guolo}, {Lu}, {Rest}, {Ricci}, \& {van Velzen}}]{masterson2025}
{Masterson}, M., {De}, K., {Panagiotou}, C., {et~al.} 2025, arXiv e-prints,
  arXiv:2503.08647

\bibitem[{{Mathis} {et~al.}(1977){Mathis}, {Rumpl}, \&
  {Nordsieck}}]{Mathis1977}
{Mathis}, J.~S., {Rumpl}, W., \& {Nordsieck}, K.~H. 1977, \apj, 217, 425

\bibitem[{{Mattila} {et~al.}(2018){Mattila}, {P{\'e}rez-Torres}, {Efstathiou},
  {Mimica}, {Fraser}, {Kankare}, {Alberdi}, {Aloy}, {Heikkil{\"a}}, {Jonker},
  {Lundqvist}, {Mart{\'\i}-Vidal}, {Meikle}, {Romero-Ca{\~n}izales}, {Smartt},
  {Tsygankov}, {Varenius}, {Alonso-Herrero}, {Bondi}, {Fransson},
  {Herrero-Illana}, {Kangas}, {Kotak}, {Ram{\'\i}rez-Olivencia},
  {V{\"a}is{\"a}nen}, {Beswick}, {Clements}, {Greimel}, {Harmanen},
  {Kotilainen}, {Nandra}, {Reynolds}, {Ryder}, {Walton}, {Wiik}, \&
  {{\"O}stlin}}]{mattila2018}
{Mattila}, S., {P{\'e}rez-Torres}, M., {Efstathiou}, A., {et~al.} 2018,
  Science, 361, 482

\bibitem[{{Meisner} {et~al.}(2021){Meisner}, {Lang}, {Schlafly}, \&
  {Schlegel}}]{Meisner2021}
{Meisner}, A.~M., {Lang}, D., {Schlafly}, E.~F., \& {Schlegel}, D.~J. 2021,
  Research Notes of the American Astronomical Society, 5, 168

\bibitem[{{Meisner} {et~al.}(2017){Meisner}, {Lang}, \&
  {Schlegel}}]{Meisner2017}
{Meisner}, A.~M., {Lang}, D., \& {Schlegel}, D.~J. 2017, \aj, 153, 38

\bibitem[{{Meisner} {et~al.}(2018){Meisner}, {Lang}, \&
  {Schlegel}}]{Meisner2018}
{Meisner}, A.~M., {Lang}, D., \& {Schlegel}, D.~J. 2018, \aj, 156, 69

\bibitem[{{Metzger}(2022)}]{Metzger2022}
{Metzger}, B.~D. 2022, \apjl, 937, L12

\bibitem[{{Mezger} {et~al.}(1996){Mezger}, {Duschl}, \& {Zylka}}]{Mezger1996}
{Mezger}, P.~G., {Duschl}, W.~J., \& {Zylka}, R. 1996, \aapr, 7, 289

\bibitem[{{Mummery} {et~al.}(2025){Mummery}, {Guolo}, {Matthews}, {Newsome},
  {Lintott}, \& {Keel}}]{Mummery2025}
{Mummery}, A., {Guolo}, M., {Matthews}, J., {et~al.} 2025, arXiv e-prints,
  arXiv:2503.14163

\bibitem[{{Nagao} {et~al.}(2017){Nagao}, {Maeda}, \& {Yamanaka}}]{Nagao2017}
{Nagao}, T., {Maeda}, K., \& {Yamanaka}, M. 2017, \apj, 835, 143

\bibitem[{{Parkinson} {et~al.}(2024){Parkinson}, {Knigge}, {Dai}, {Thomsen},
  {Matthews}, \& {Long}}]{Parkinson2024}
{Parkinson}, E.~J., {Knigge}, C., {Dai}, L., {et~al.} 2024, arXiv e-prints,
  arXiv:2408.16371

\bibitem[{{Phinney}(1989)}]{Phinney1989}
{Phinney}, E.~S. 1989, in The Center of the Galaxy, ed. M.~{Morris}, Vol. 136,
  543

\bibitem[{{Piran} {et~al.}(2015){Piran}, {Svirski}, {Krolik}, {Cheng}, \&
  {Shiokawa}}]{Piran2015}
{Piran}, T., {Svirski}, G., {Krolik}, J., {Cheng}, R.~M., \& {Shiokawa}, H.
  2015, \apj, 806, 164

\bibitem[{{Rees}(1988)}]{Rees1988}
{Rees}, M.~J. 1988, \nat, 333, 523

\bibitem[{{Reynolds} {et~al.}(2022){Reynolds}, {Mattila}, {Efstathiou},
  {Kankare}, {Kool}, {Ryder}, {Pe{\~n}a-Mo{\~n}ino}, \&
  {P{\'e}rez-Torres}}]{reynolds2022}
{Reynolds}, T.~M., {Mattila}, S., {Efstathiou}, A., {et~al.} 2022, \aap, 664,
  A158

\bibitem[{{Rossi} {et~al.}(2021){Rossi}, {Stone}, {Law-Smith}, {Macleod},
  {Lodato}, {Dai}, \& {Mandel}}]{Rossi2021}
{Rossi}, E.~M., {Stone}, N.~C., {Law-Smith}, J.~A.~P., {et~al.} 2021, \ssr,
  217, 40

\bibitem[{{Roth} {et~al.}(2016){Roth}, {Kasen}, {Guillochon}, \&
  {Ramirez-Ruiz}}]{Roth2016}
{Roth}, N., {Kasen}, D., {Guillochon}, J., \& {Ramirez-Ruiz}, E. 2016, \apj,
  827, 3

\bibitem[{{Roth} {et~al.}(2020){Roth}, {Rossi}, {Krolik}, {Piran}, {Mockler},
  \& {Kasen}}]{Roth2020}
{Roth}, N., {Rossi}, E.~M., {Krolik}, J., {et~al.} 2020, \ssr, 216, 114

\bibitem[{{Saxton} {et~al.}(2020){Saxton}, {Komossa}, {Auchettl}, \&
  {Jonker}}]{Saxton2020}
{Saxton}, R., {Komossa}, S., {Auchettl}, K., \& {Jonker}, P.~G. 2020, \ssr,
  216, 85

\bibitem[{{Schlafly} {et~al.}(2019){Schlafly}, {Meisner}, \&
  {Green}}]{Schlafly2019}
{Schlafly}, E.~F., {Meisner}, A.~M., \& {Green}, G.~M. 2019, \apjs, 240, 30

\bibitem[{{Skrutskie} {et~al.}(2006){Skrutskie}, {Cutri}, {Stiening},
  {Weinberg}, {Schneider}, {Carpenter}, {Beichman}, {Capps}, {Chester},
  {Elias}, {Huchra}, {Liebert}, {Lonsdale}, {Monet}, {Price}, {Seitzer},
  {Jarrett}, {Kirkpatrick}, {Gizis}, {Howard}, {Evans}, {Fowler}, {Fullmer},
  {Hurt}, {Light}, {Kopan}, {Marsh}, {McCallon}, {Tam}, {Van Dyk}, \&
  {Wheelock}}]{Skrutskie2006}
{Skrutskie}, M.~F., {Cutri}, R.~M., {Stiening}, R., {et~al.} 2006, \aj, 131,
  1163

\bibitem[{{Smartt} {et~al.}(2015){Smartt}, {Valenti}, {Fraser}, {Inserra},
  {Young}, {Sullivan}, {Pastorello}, {Benetti}, {Gal-Yam}, {Knapic},
  {Molinaro}, {Smareglia}, {Smith}, {Taubenberger}, {Yaron}, {Anderson},
  {Ashall}, {Balland}, {Baltay}, {Barbarino}, {Bauer}, {Baumont}, {Bersier},
  {Blagorodnova}, {Bongard}, {Botticella}, {Bufano}, {Bulla}, {Cappellaro},
  {Campbell}, {Cellier-Holzem}, {Chen}, {Childress}, {Clocchiatti},
  {Contreras}, {Dall'Ora}, {Danziger}, {de Jaeger}, {De Cia}, {Della Valle},
  {Dennefeld}, {Elias-Rosa}, {Elman}, {Feindt}, {Fleury}, {Gall},
  {Gonzalez-Gaitan}, {Galbany}, {Morales Garoffolo}, {Greggio}, {Guillou},
  {Hachinger}, {Hadjiyska}, {Hage}, {Hillebrandt}, {Hodgkin}, {Hsiao}, {James},
  {Jerkstrand}, {Kangas}, {Kankare}, {Kotak}, {Kromer}, {Kuncarayakti},
  {Leloudas}, {Lundqvist}, {Lyman}, {Hook}, {Maguire}, {Manulis}, {Margheim},
  {Mattila}, {Maund}, {Mazzali}, {McCrum}, {McKinnon}, {Moreno-Raya},
  {Nicholl}, {Nugent}, {Pain}, {Pignata}, {Phillips}, {Polshaw}, {Pumo},
  {Rabinowitz}, {Reilly}, {Romero-Ca{\~n}izales}, {Scalzo}, {Schmidt},
  {Schulze}, {Sim}, {Sollerman}, {Taddia}, {Tartaglia}, {Terreran},
  {Tomasella}, {Turatto}, {Walker}, {Walton}, {Wyrzykowski}, {Yuan}, \&
  {Zampieri}}]{Smartt2015}
{Smartt}, S.~J., {Valenti}, S., {Fraser}, M., {et~al.} 2015, \aap, 579, A40

\bibitem[{{Thomsen} {et~al.}(2022){Thomsen}, {Kwan}, {Dai}, {Wu}, {Roth}, \&
  {Ramirez-Ruiz}}]{Thomsen2022}
{Thomsen}, L.~L., {Kwan}, T.~M., {Dai}, L., {et~al.} 2022, \apjl, 937, L28

\bibitem[{{Ulmer}(1999)}]{Ulmer1999}
{Ulmer}, A. 1999, \apj, 514, 180

\bibitem[{{van Velzen} {et~al.}(2021){van Velzen}, {Gezari}, {Hammerstein},
  {Roth}, {Frederick}, {Ward}, {Hung}, {Cenko}, {Stein}, {Perley}, {Taggart},
  {Foley}, {Sollerman}, {Blagorodnova}, {Andreoni}, {Bellm}, {Brinnel}, {De},
  {Dekany}, {Feeney}, {Fremling}, {Giomi}, {Golkhou}, {Graham}, {Ho},
  {Kasliwal}, {Kilpatrick}, {Kulkarni}, {Kupfer}, {Laher}, {Mahabal}, {Masci},
  {Miller}, {Nordin}, {Riddle}, {Rusholme}, {van Santen}, {Sharma}, {Shupe}, \&
  {Soumagnac}}]{vanVelzen2021}
{van Velzen}, S., {Gezari}, S., {Hammerstein}, E., {et~al.} 2021, \apj, 908, 4

\bibitem[{{van Velzen} {et~al.}(2016){van Velzen}, {Mendez}, {Krolik}, \&
  {Gorjian}}]{vanVelzen2016}
{van Velzen}, S., {Mendez}, A.~J., {Krolik}, J.~H., \& {Gorjian}, V. 2016,
  \apj, 829, 19

\bibitem[{{Wevers}(2020)}]{Wevers2020}
{Wevers}, T. 2020, \mnras, 497, L1

\bibitem[{{Wevers} \& {French}(2024)}]{wevers24}
{Wevers}, T. \& {French}, K.~D. 2024, \apjl, 969, L17

\bibitem[{{Wright} \& {Barlow}(1975)}]{Wright1975}
{Wright}, A.~E. \& {Barlow}, M.~J. 1975, \mnras, 170, 41

\bibitem[{{Yao} {et~al.}(2023){Yao}, {Ravi}, {Gezari}, {van Velzen}, {Lu},
  {Schulze}, {Somalwar}, {Kulkarni}, {Hammerstein}, {Nicholl}, {Graham},
  {Perley}, {Cenko}, {Stein}, {Ricarte}, {Chadayammuri}, {Quataert}, {Bellm},
  {Bloom}, {Dekany}, {Drake}, {Groom}, {Mahabal}, {Prince}, {Riddle},
  {Rusholme}, {Sharma}, {Sollerman}, \& {Yan}}]{Yao2023}
{Yao}, Y., {Ravi}, V., {Gezari}, S., {et~al.} 2023, \apjl, 955, L6

\bibitem[{{Yuan} {et~al.}(2013){Yuan}, {Liu}, \& {Xiang}}]{Yuan2013}
{Yuan}, H.~B., {Liu}, X.~W., \& {Xiang}, M.~S. 2013, \mnras, 430, 2188

\bibitem[{{Zhuang} {et~al.}(2025){Zhuang}, {Shen}, {Mou}, \& {Lu}}]{Zhuang2025}
{Zhuang}, J., {Shen}, R.-F., {Mou}, G., \& {Lu}, W. 2025, \apj, 979, 109

\end{thebibliography}

 \begin{appendix}

\section{The contribution from free-free opacity in our modelling}
\label{appendix_sec:free-free}
\begin{figure}
   \centering
   \includegraphics[width=0.49\textwidth]{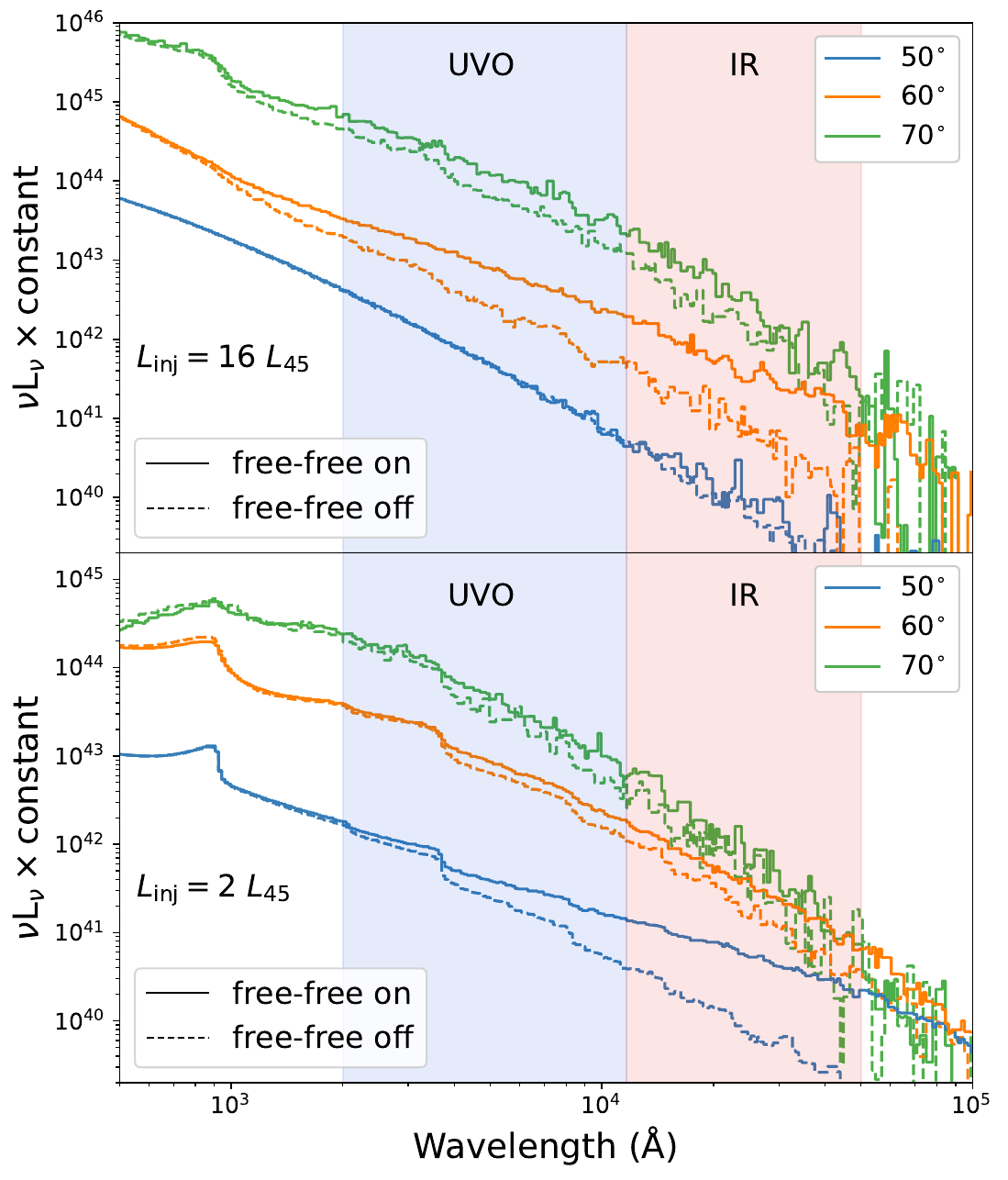} \newline
      \caption{Simulated escaping spectra from our reprocessing model, with free-free opacity and emission on and off for the spectra drawn with a solid or dashed line respectively. All spectra have \macc~=24\medd~ and $\rho_{\rm scale}=1$. The top and bottom panels have $L_{\rm inj}=16\times10^{45}$ and $2\times10^{45}$ erg s$^{-1}$ respectively. The 50\textdegree~and 70\textdegree~spectra have been scaled by a factor of 0.1 and 5 respectively, to improve visual clarity. The spectral regions covered by UVO and IR observations are indicated by the blue and red shading respectively.
      }
\label{fig:free-free_on-off}
\end{figure}

The NIR slope for AT~2019azh is shallower than the Rayleigh-Jeans part of a blackbody spectrum, and to reproduce this, we require a strong contribution from free-free emission. The strength of the free-free opacity is closely tied to the ionisation state, which decreases with increasing density (i.e. increasing $\rho_{\rm scale}$ or $\theta$), and increases with $L_{\rm inj}$. In Fig. \ref{fig:free-free_on-off}, we show simulated escaping spectra from our reprocessing model with inclinations of $\theta=50^{\circ},60^{\circ},70^{\circ}$ for a large and small value of $L_{\rm inj}$. We additionally show the same model after turning off the contribution from free-free opacity and emission, to illustrate its relative importance for different parameters and at different wavelengths.

In high ionisation states, for example, for a lower inclination angle of $\theta=50$\textdegree~ and a large $L_{\rm inj}=16~L_{45}$, the gas is highly ionised and hot. The contribution from free-free emission in these cases is weak, as it scales with temperature as $T^{-7/2}$ according to Kramers' law. Therefore, all opacity and emission terms are weak, and electron scattering is dominant. The injected radiation interacts with the outflowing wind, and bulk scattering (adiabatic work) stretches the initial blackbody spectrum, resulting in a blackbody like slope in the IR. By contrast, in low ionisation states, for example, a larger inclination angle of $\theta=70$\textdegree~ and a low $L_{\rm inj}=2~L_{45}$, bound-free processes dominate, free-free emission makes only a minor contribution, and the spectral is again blackbody-like. For the $\theta=70$\textdegree~case, the free-free emission is minimal due to a low level of ionisation, even when $L_{\rm inj}$ is large, so that there can never be a large IR excess. When the ionisation state is intermediate between the two extremes, for example, for $\theta=60$\textdegree~and $L_{\rm inj}=16~L_{45}$ or $\theta=50$\textdegree~and $L_{\rm inj}=2~L_{45}$ we can see that the free-free emission contribution can be very significant.
 
\section{The IR echo model for the early data}

\begin{figure}
   \centering
   \includegraphics[width=0.49\textwidth]{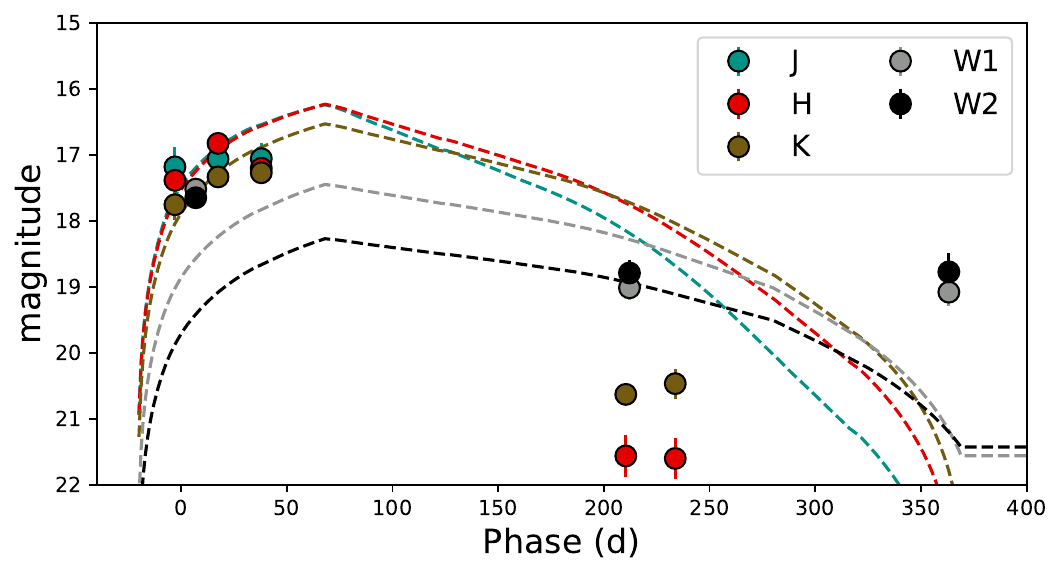} \newline
   \includegraphics[width=0.49\textwidth]{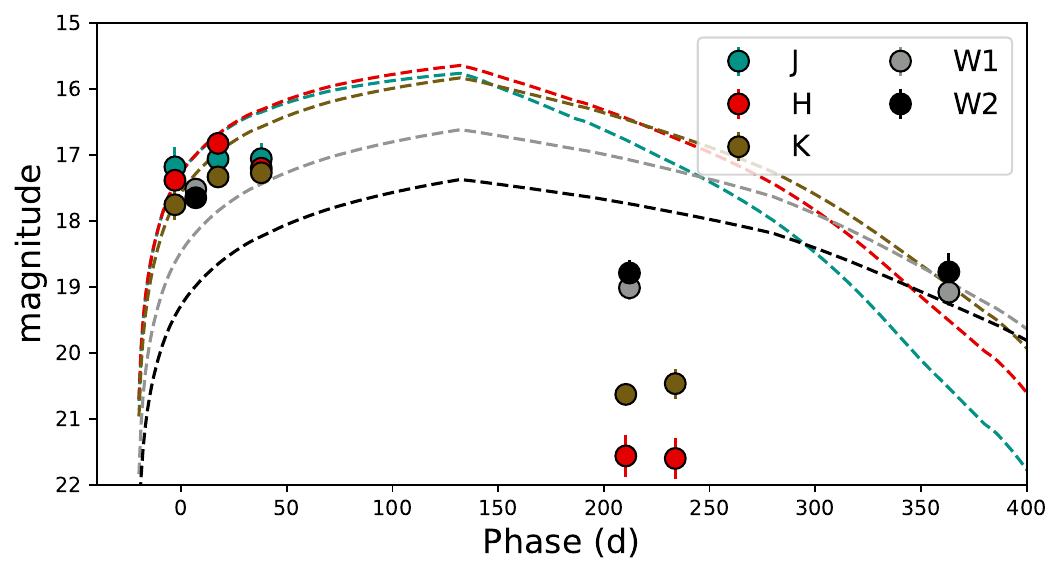} \newline
      \caption{
      \textbf{Top}: IR echo model for a spherical shell of graphite dust with a single size composition of 0.1~$\mu$m grains. The dust shell has radius 0.037~pc, corresponding to the sublimation radius inferred from the TDE luminosity. \textbf{Bottom}: As top panel, for a multiple size composition with radius 0.064~pc.
      }
\label{fig:IR_echo_revp}
\end{figure}

We consider the case in which the TDE evaporates dust out to a certain radius ($R_{\text{sub}}$), and that there is a spatially-thin spherical shell of dust at that radius. The IR echo signal will be dominated by the emission arising at the inner part of any dust distribution, where the dust is hottest, the density highest and the IR echo has a shorter duration, due to shorter light travel times, so the spatially-thin shell is a reasonable approproximation. The results of our temperature fitting in Sect. \ref{subsec:dust_fitting} indicated that if the IR echo model is applicable, the temperature of the dust at early times is $\sim$2100~K. This is approximately the sublimation temperature for graphite, and is too high for the survival of silicates, so we assume that the dust is pure graphite and consider both a single grain size distribution of 0.1~$\mu$m grains and a multiple grain size distribution with minimum size 0.005~$\mu$m, number density of grains following $a^{-3.5}$, and maximum size 0.1~$\mu$m. Assuming based on the SED fitting that the dust sublimation temperature is 2100~K, we make use of Eq. (\ref{equation:model_temp}) to find that the resulting evaporation radii are $R_{\text{evap,single}}$ = 0.037 pc and $R_{\text{evap,multiple}}$ = 0.064 pc for the single and multiple grain size distributions, respectively. These are lower limits for the sublimation radius, as any additional extreme-UV or X-ray emission from the TDE compared to our assumed blackbody spectrum will contribute to evaporating dust to a larger radius. We then use Eqn. \ref{equation:model_time} to derive the IR echo behaviour for these radii and dust compositions. We show the result for the single and multiple grain size distributions in Fig. \ref{fig:IR_echo_revp}. 

The resulting IR echoes do not fit the early time emission well. The time evolution in the NIR is not well reproduced, although there are significant systematic uncertainties introduced by subtracting the estimated TDE contribution in the IR, and the MIR data are much brighter than predicted. Furthermore, at later times, nearly all the data is fainter than predicted. In order to reconcile these observations with an echo model, one would require a clump of dust close to the line of sight with small enough opening angle that it does not contribute significantly to the observations later than 200~d, as well as another more distant clump to reproduce the MIR at early times, and then clumps such as those we find in Sect. \ref{subsubsec:blob} for the late time data. Although this is feasible, such a model has too many free parameters to sensible constrain with our data.

\section{Additional figures}

\begin{figure}
   \centering
   \includegraphics[width=0.5\textwidth]{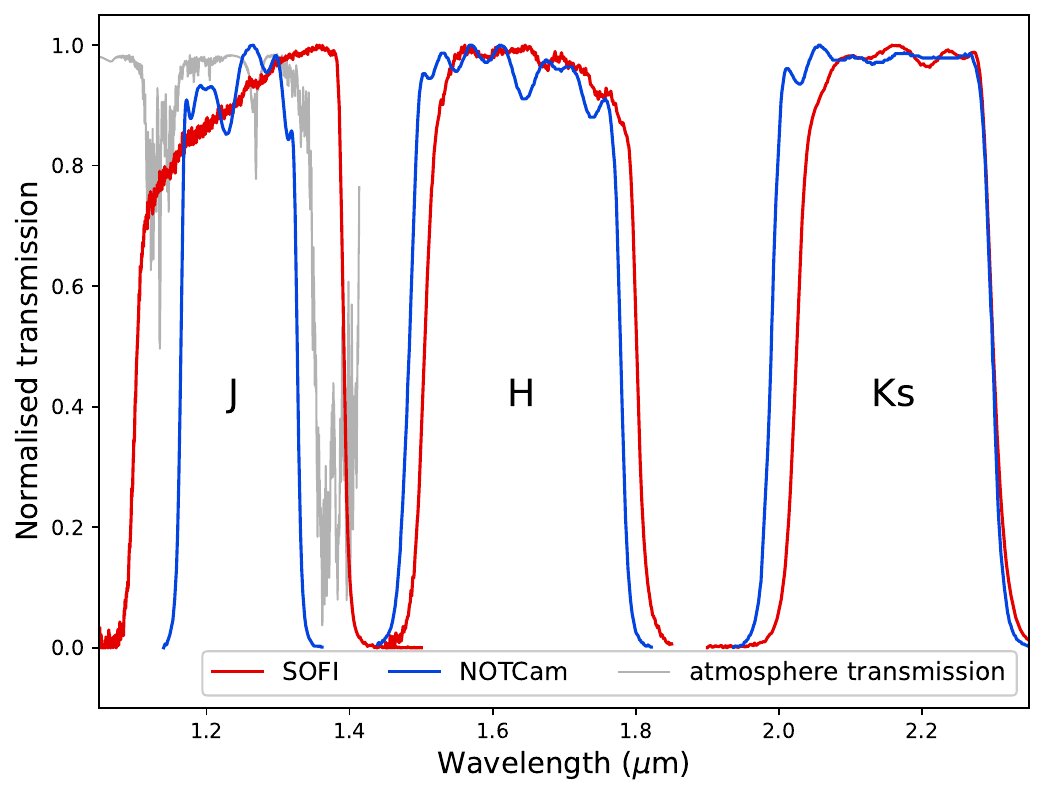}
      \caption{Normalised $JHK$ transmission curves for SOFI and NOTCam. The atmospheric transmission in the region around the $J$ filter is also shown.
      }
\label{fig:filter_functions}
\end{figure}

\begin{figure}
   \centering
   \includegraphics[width=0.5\textwidth]{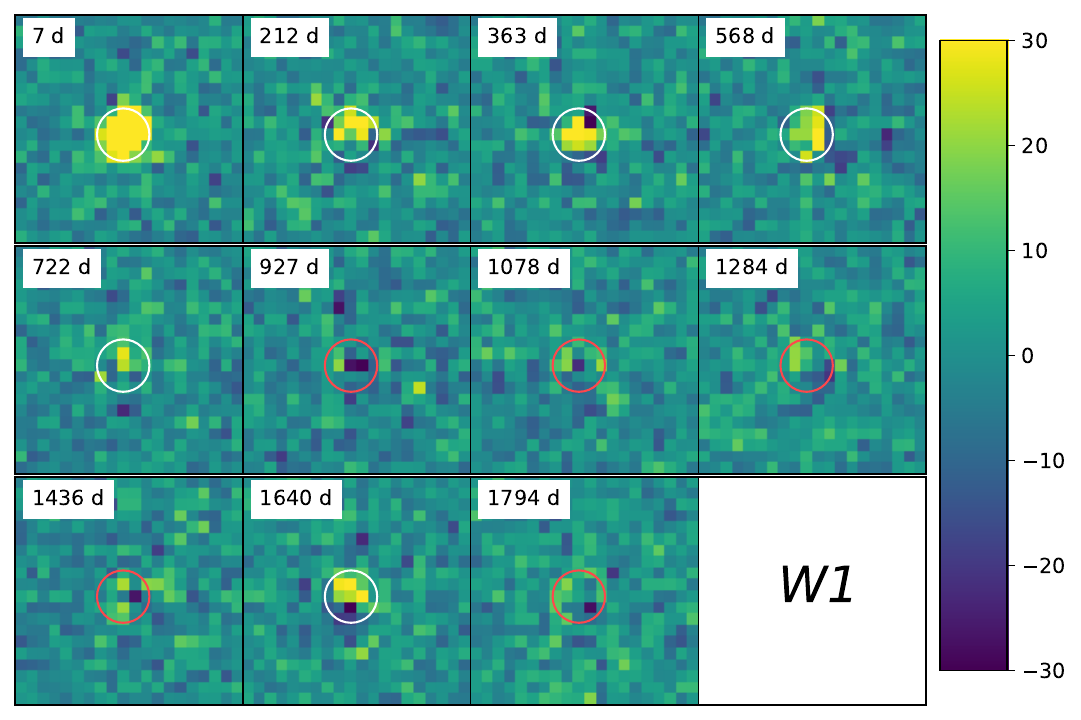} \\
   \includegraphics[width=0.5\textwidth]{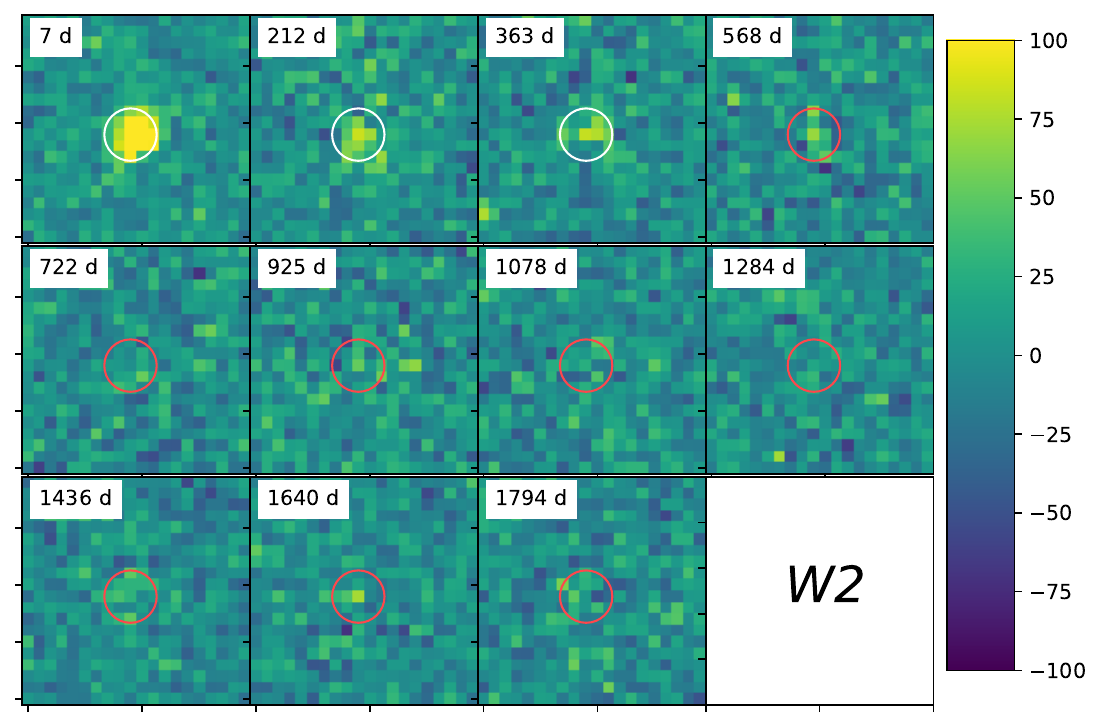}
      \caption{Subtraction residuals after template subtraction for the unWISE data, with the $W1$ and $W2$ band in the upper and lower panels respectively. A linear scaling that shows positive and negative values symmetrically around zero is used to clearly show dipole-like subtraction residuals. White and red circles indicate detections and non-detections, respectively. The radius of the circle corresponds to the FWHM of the image PSF.}
\label{fig:NEOWISE_subtractions}
\end{figure}

\begin{figure}
   \centering
   \includegraphics[width=0.5\textwidth]{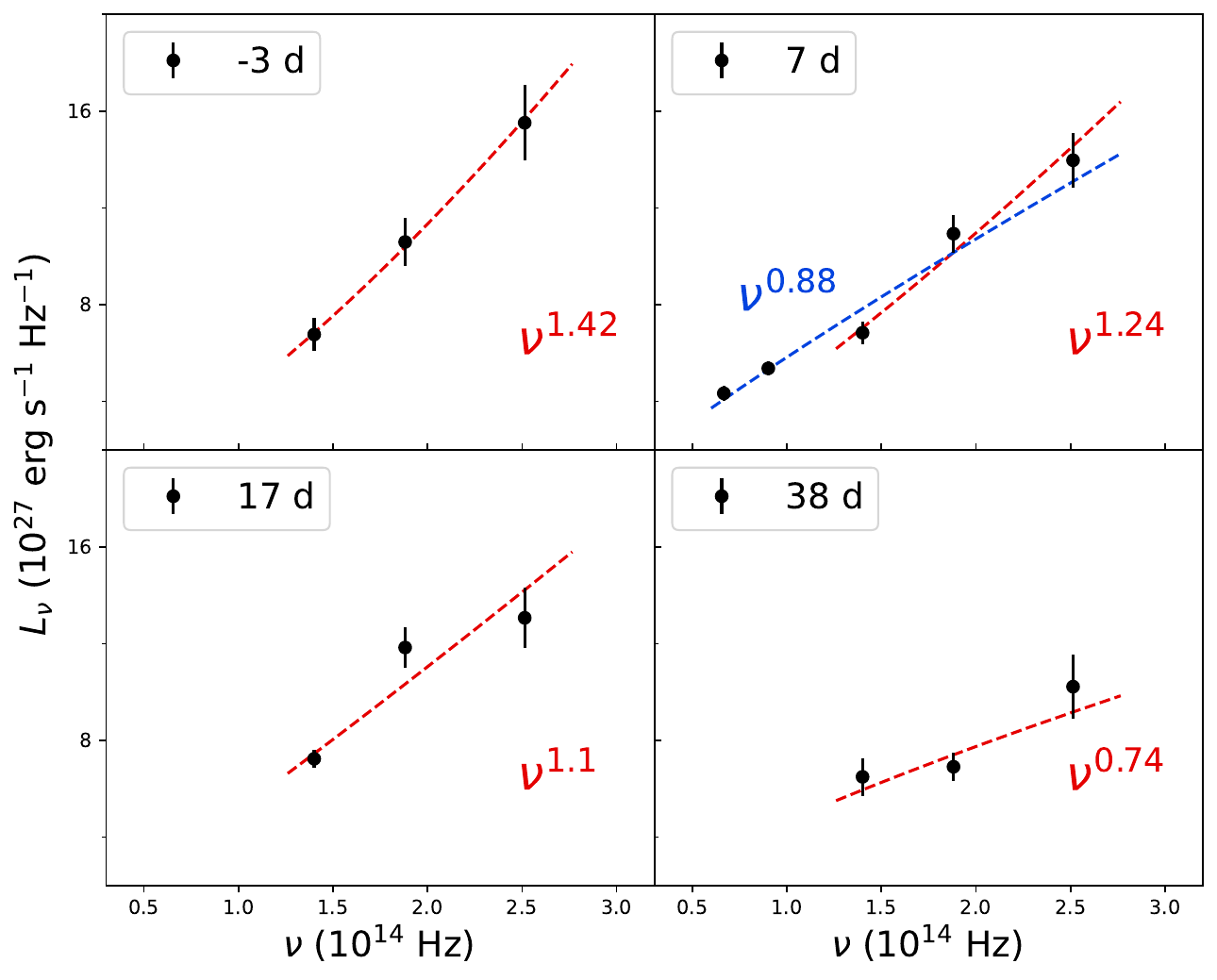}
      \caption{Power law fits to the NIR and MIR SEDs of AT~2019azh. The power laws shown in red fit only the NIR data, while those shown in blue fit both the NIR and MIR data. Uncertainties are listed in Tab. \ref{tab:power_law_fitting}.
      }
\label{fig:power_law_fit}
\end{figure}

\begin{figure*}
    
   \centering
   \includegraphics[width=1\textwidth]{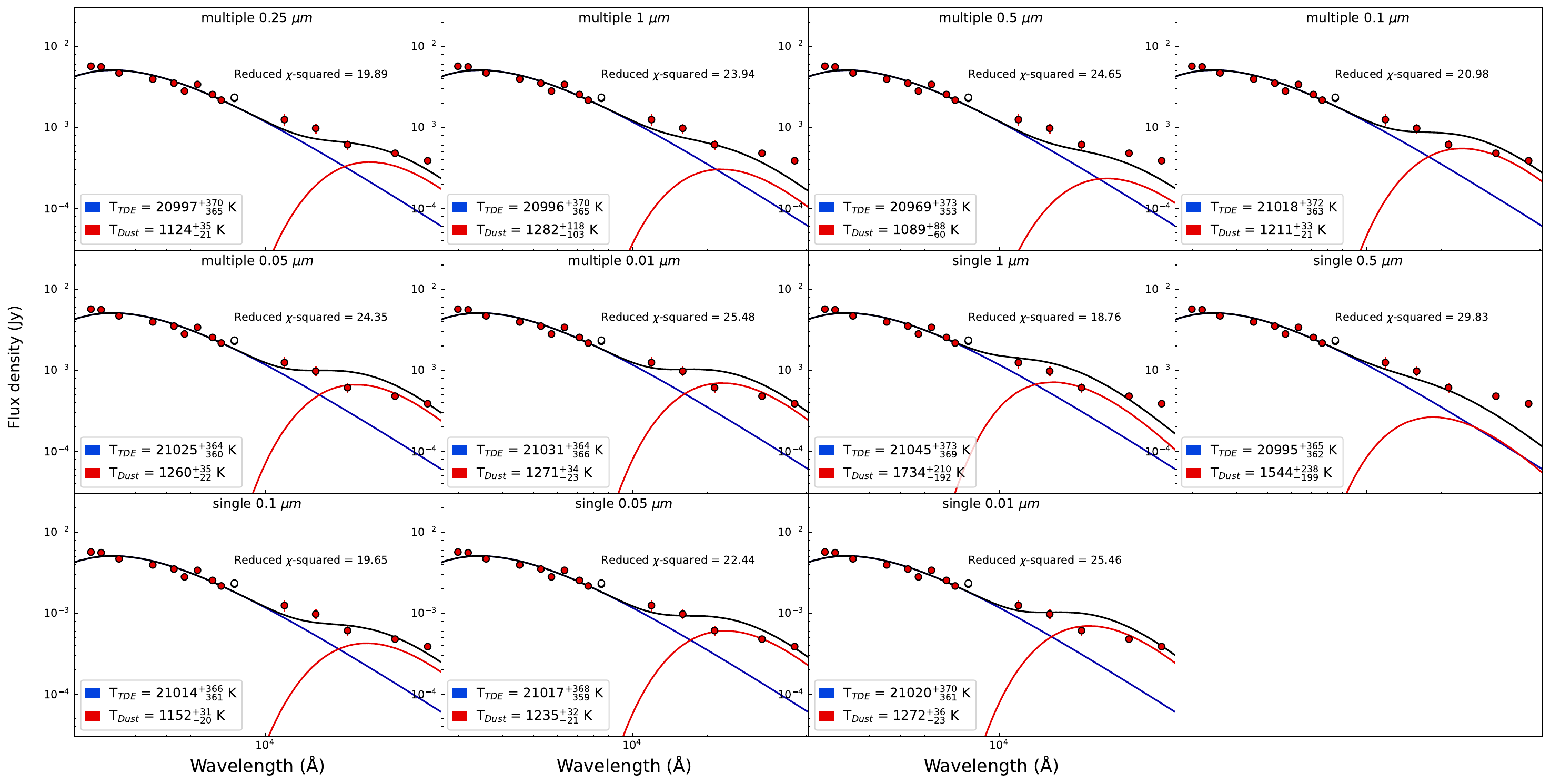}
      \caption{SED fitting with two blackbodies for the observed UVO+IR SEDs. The fitting was performed with a modified blackbody for the dust. The dust composed of graphite with a variety of grain sizes, as described in Sect. \ref{subsec:dust_fitting}, including both single size and multiple size distributions. The $i$ band was excluded from the fit.
      }
\label{fig:SED_fitting_many_comps}
\end{figure*}

%\begin{figure}
%   \centering
%   \includegraphics[width=0.5\textwidth]{figures/19azh_LC_peak.png}
%      \caption{Optical and IR data of AT 2019azh around peak, before the first solar conjunction.
%      }
%\label{fig:LC_at_peak}
%\end{figure}

%\begin{figure}
%    \centering
%   \includegraphics[width=0.49\textwidth]{figures/IR_echo_models/IR_echo_model_graphite_multiple_0_1_0_3pc.png}
%    \caption{Spherical shell model ignoring the colours - 0.3 pc shell can reproduce more or less. Adding thickness and bringing the radius in a bit more would be a better fit. But colours are wildly inconsistent.}
%    \label{fig:W1_optimised_model}
%\end{figure}

\section{Tables}

\begin{table*}
\caption{Photometry of AT~2019azh. All magnitudes are given in the AB system and have been corrected for Milky Way extinction. Phases are in rest frame days given relative to $g$-band peak. Limits listed are 3$\sigma$ upper limits.}
\centering
%\begin{adjustbox}{width=0.5\textwidth}
\begin{tabular}{c c c c c c c c} \hline\hline % c|c|
Phase & Date & MJD & Telescope & Instrument & Band & Mag & Uncertainty  \\ 
%& Transient Flux density & Transient Luminosity \\ 
%(1)&(2)&(3)&(4)&(5)&(6) \\ \hline

\hline
-2.98 & 2019-03-20 & 58562.11 & NTT & SOFI & J & 16.06 & 0.11 \\
17.41 & 2019-04-09 & 58582.96 & NOT & NOTCam & J & 16.25 & 0.10 \\
37.87 & 2019-04-30 & 58603.87 & NOT & NOTCam & J & 16.52 & 0.14 \\
210.32 & 2019-10-23 & 58780.16 & NOT & NOTCam & J & 20.43 & 0.10 \\
233.73 & 2019-11-17 & 58804.08 & NOT & NOTCam & J & 20.38 & 0.17 \\

-2.98 & 2019-03-20 & 58562.11 & NTT & SOFI & H & 16.48 & 0.10  \\
17.42 & 2019-04-09 & 58582.97 & NOT & NOTCam & H & 16.36 & 0.08 \\
37.87  & 2019-04-30 & 58603.87 & NOT & NOTCam & H & 16.94 & 0.09 \\
210.33  & 2019-10-23 & 58780.17 & NOT & NOTCam & H & 20.38 & 0.11 \\
233.74 & 2019-11-17 & 58804.09 & NOT & NOTCam & H & 20.37 & 0.10 \\

-2.98 & 2019-03-20 & 58562.12 & NTT & SOFI & K & 16.97 & 0.11 \\
17.42 & 2019-04-09 & 58582.97 & NOT & NOTCam & K & 16.89 & 0.06 \\
37.88  & 2019-04-30 & 58603.88 & NOT & NOTCam & K & 17.01 & 0.13 \\
210.34  & 2019-10-23 & 58780.18 & NOT & NOTCam & K & 20.20 & 0.11 \\
233.75 & 2019-11-17 & 58804.10 & NOT & NOTCam & K & 20.08 & 0.16 \\

6.88 & 2019-03-30 & 58572.20 & WISE & WISE & W1 & 17.22 & 0.06 \\ 
212.08 & 2019-10-25 & 58781.95 & WISE & WISE & W1 & 18.96 & 0.17 \\ 
363.18 & 2020-03-28 & 58936.41 & WISE & WISE & W1 & 19.17 & 0.20 \\ 
568.37 & 2020-10-24 & 59146.16 & WISE & WISE & W1 & 19.41 & 0.26 \\ 
722.29 & 2021-03-30 & 59303.51 & WISE & WISE & W1 & 19.72 & 0.21 \\ 
926.73 & 2021-10-25 & 59512.48 & WISE & WISE & W1 & >19.90 & - \\ 
1078.48 & 2022-03-29 & 59667.60 & WISE & WISE & W1 & >19.81 & - \\ 
1283.80 & 2022-10-25 & 59877.50 & WISE & WISE & W1 & >19.96 & - \\ 
1436.25 & 2022-03-30 & 60033.33 & WISE & WISE & W1 & >20.03 & - \\
1640.05 & 2023-10-24 & 60241.66 & WISE & WISE & W1 & 19.38 & 0.35 \\ 
1793.99 & 2024-03-30 & 60399.02 & WISE & WISE & W1 & >20.07 & - \\ 

6.88 & 2019-03-30 & 58572.20 & WISE & WISE & W2 & 17.45 & 0.08 \\ 
212.08 & 2019-10-25 & 58781.95 & WISE & WISE & W2 & 18.77 & 0.19 \\ 
363.18 & 2020-03-28 & 58936.41 & WISE & WISE & W2 & 18.77 & 0.29 \\ 
568.37 & 2020-10-24 & 59146.16 & WISE & WISE & W2 & >19.17 & - \\ 
722.29 & 2021-03-30 & 59303.51 & WISE & WISE & W2 & >19.03 & - \\ 
924.68 & 2021-10-23 & 59510.39 & WISE & WISE & W2 & >19.16 & - \\ 
1078.48 & 2022-03-29 & 59667.60 & WISE & WISE & W2 & >19.08 & - \\ 
1283.81 & 2022-10-25 & 59877.50 & WISE & WISE & W2 & >18.95 & - \\ 
1436.25 & 2022-03-30 & 60033.33 & WISE & WISE & W2 & >19.20 & - \\ 
1640.06 & 2023-10-24 & 60241.66 & WISE & WISE & W2 & >19.20 & - \\ 
1794.00 & 2024-03-30 & 60399.02 & WISE & WISE & W2 & >19.01 & - \\ 

\end{tabular}
%\end{adjustbox}
%; (7) the flux density after correcting for the host contribution; and (8) the corresponding luminosity.}
\label{tab:IR_photometry}
\end{table*}

 \end{appendix}

\end{document}